\documentstyle[11pt,aaspp4]{article}

\newcommand\nion[2]{#1\,\lowercase{{\sc #2}}}
\newcommand\wave[1]{\mbox{$\lambda$#1\,\AA}}
\def\kmsec{\mbox{km~s$^{\rm -1}$}}
\def\teff{\mbox{T$_{\rm eff}$}}
\def\logg{\mbox{$log$~g}}
\def\vmicro{\mbox{v$_{\rm t}$}}
\def\BmV0{\mbox{$(B-V)_{\rm o}$}}
\def\VmK0{\mbox{$(V-K)_{\rm o}$}}
\def\MV0{\mbox{$M_{\rm V}_{\rm o}$}}

\def\carbiso{\mbox{${\rm ^{12}C/^{13}C}$}}
\def\etal{\mbox{{\it et al.}}}
\def\eg{\mbox{{\it e.g.}}}
\def\ie{\mbox{{\it i.e.}}}
 
\lefthead{}
\righthead{}
\begin{document}

\title{Star-to-Star Abundance Variations among Bright Giants in the
Mildly Metal-Poor Globular Cluster M4}

\author{
Inese I. Ivans\altaffilmark{1}, 
Christopher Sneden\altaffilmark{1}, 
Robert P. Kraft\altaffilmark{2},
Nicholas B. Suntzeff\altaffilmark{3},
Verne V. Smith\altaffilmark{4,5},
G. Edward Langer\altaffilmark{6,7},
Jon P. Fulbright\altaffilmark{2}
}

\altaffiltext{1}{Department of Astronomy and McDonald Observatory, 
University of Texas, Austin, TX 78712; iivans@astro.as.utexas.edu, 
chris@verdi.as.utexas.edu}

\altaffiltext{2}{UCO/Lick Observatory, Board of Studies in Astronomy and
Astrophysics, University of California, Santa Cruz, CA  95064;
kraft@ucolick.org}

\altaffiltext{3}{Cerro Tololo Inter-American Observatory, National
Optical Astronomy Observatories, which is operated by the Association of 
Universities for Research in Astronomy, Inc., (AURA), under cooperative 
agreement with the National Science Foundation.  La Serena, Chile; 
nsuntzeff@noao.edu}

\altaffiltext{4}{Department of Physics, University of Texas at El Paso,
500 West University, El Paso, TX 79968-0515; verne@balmer.physics.utep.edu}

\altaffiltext{5}{Visiting Astronomer, Cerro Tololo Inter-American 
Observatory, National Optical Astronomy Observatories, which is operated 
by the Association of Universities for Research in Astronomy, Inc., (AURA), 
under cooperative agreement with the National Science Foundation.}

\altaffiltext{6}{Physics Department, Colorado College, Colorado Springs, CO
80903; elanger@academic.cc.colorado.edu}

\altaffiltext{7}{Deceased 1999 February 16.}
 
\vskip .5truein
\begin{center}
Submitted to {\it The Astronomical Journal}
\end{center}

\begin{abstract}

We present a chemical composition analysis of 36 giants in the nearby 
mildy metal-poor ($<$[Fe/H]$>$ = --1.18) ``CN-bimodal" globular cluster 
M4. The stars were observed at the Lick \& McDonald Observatories using 
high resolution \'echelle spectrographs and at CTIO using the multi-object 
spectrometer. Confronted with a cluster having interstellar extinction 
that is large and variable across the cluster face, we combined 
traditional spectroscopic abundance methods with modifications to the 
line-depth ratio technique pioneered by Gray (1994\markcite{Gr94}) to
determine the atmospheric parameters of our stars.  We derive a 
total-to-selective extinction ratio of 3.4~$\pm$~0.4 and an average 
$<$E(B--V)$>$ reddening of 0.33~$\pm$~0.01 which is significantly lower 
than that estimated by using the dust maps made by Schlegel \etal\ 
(1998\markcite{SFD98}).

We determine abundance ratios typical of halo field and cluster stars for 
scandium, titanium, vanadium, nickel, and europium with star-to-star 
variations in these elements of $<$~$\pm$0.1.  Silicon, aluminum, barium,
and lanthanum are overabundant with respect to what is seen in other
globular clusters of similar metallicity.  These overabundances confirm 
the results of an earlier study based on a much smaller sample of M4 
giants (Brown \& Wallerstein 1992\markcite{BW92}).

Superimposed on the primordial abundance distribution is evidence for the 
existence of proton-capture synthesis of carbon, oxygen, neon, and 
magnesium. We recover some of the C, N, O, Na, Mg, and Al abundance swings 
and correlations found in other more metal-poor globular clusters but the 
range of variation is muted.  In the case of Mg and Al, this is compatible 
with the idea that the Al enhancements are derived from the destruction of 
$^{25,26}$Mg, not $^{24}$Mg. We determine that the C+N+O abundance sum is 
constant to within the observational errors, and agrees with the C+N+O 
total that might be expected for M4 stars at birth.

The AGB stars in M4 have C,N,O abundances that show less evidence for 
proton-capture nucleosynthesis than is found in the less-evolved stars of 
the RGB.  Deeply-mixed stars of the RGB, subsequent to the helium core 
flash, might take up residence on the blue end of the HB, and thus fail to 
evolve back to the AGB but reasons for skepticism concerning this scenario 
are noted. 

\end{abstract}

\keywords{globular clusters: individual (NGC~6121) --- globular clusters: 
general --- stars: abundances --- nucleosynthesis --- stars: fundamental
parameters --- dust, extinction}

\section{Introduction}

Very large star-to-star abundance variations in the light elements C, N, 
O, Na, Mg and Al occur among the bright giants of a number of globular 
clusters.  In clusters where giant star samples have been sufficiently 
large, the abundances of O and Na are anticorrelated, as are those of Mg 
and Al.  Prime examples are found in M13 (Pilachowski \etal\ 
1996\markcite{PSK96}; Shetrone 1996a,b\markcite{Sh96a}\markcite{Sh96b},
Kraft \etal\ 1997\markcite{KSSSLP97}), $\omega$~Cen (Paltoglou \& Norris 
1989\markcite{PN89}; Norris \& Da Costa 
1995a,b\markcite{ND95a}\markcite{ND95b}), NGC~6752 (Cottrell \& Da Costa 
1981\markcite{CD81}), M15 (Sneden \etal\ 1997\markcite{SKSSLP97}), 
and most recently NGC~3201 (Gonzalez \& Wallerstein 1998\markcite{GW98}).
Recent major reviews of cluster abundance trends have been published by 
Suntzeff (1993)\markcite{Su93}, Briley \etal\ (1994)\markcite{BBHS94}, 
Kraft (1994)\markcite{Kr94}, and Wallerstein \etal\ 
(1997)\markcite{Wetal97}. 

Most investigators agree that the abundance anticorrelations arise from
proton-capture chains that convert C and O into N, Ne into Na, and Mg 
into Al in the hydrogen-burning layers of evolved cluster stars. 
Controversy has arisen, however, between proponents of what are termed 
``evolutionary'' {\it vs.} ``primordial'' scenarios.   For a brief 
review of these alternatives, we refer the reader to the introductory 
section of Sneden \etal\ (1997).  In the former picture one supposes 
that, in the low-mass red giants we presently observe, the products of 
internal proton-capture synthesis are brought to the surface by 
deep-mixing of the stellar envelope through the hydrogen-burning shell.  
Therefore, the anomalous behavior is expected to increase (on average) 
with advancing evolutionary state. In the latter scenario, one supposes 
that the proton-capture synthesis took place in a prior generation of 
more massive stars.  As a result of mass loss, these massive stars 
produced the required abundance redistribution in the primordial 
material out of which the presently observed low-mass stars were formed. 
Several lines of observational evidence now suggest that both scenarios 
may play important roles: superimposed on a primordial spread are 
additional variations resulting from deep-mixing in the stars we 
presently observe ({\it e.g.} Briley \etal\ 1994).

Stellar evolution theory (Sweigart \& Mengel 1979\markcite{SM79}) 
predicts that deep-mixing should become less efficient and possibly
cut off as metallicity increases. Observational support for this 
hypothesis was found by Sneden \etal\ (1994\markcite{SKLPS94}), who 
showed that the  C, N, O and Na abundances have little variation among 
giants in M71 ([Fe/H]~= --0.8),\footnote{We adopt the usual 
spectroscopic notations that
[A/B]~$\equiv$~{\rm log}$_{\rm 10}$(N$_{\rm A}$/N$_{\rm B}$)$_{\rm star}$~--
{\rm log}$_{\rm 10}$(N$_{\rm A}$/N$_{\rm B}$)$_{\odot}$, and that 
{\rm log}~$\epsilon$(A)~$\equiv$~{\rm log}$_{\rm 10}$(N$_{\rm A}$/N$_{\rm H}$)~+~12.0,
for elements A and B. Also, metallicity is defined as the stellar [Fe/H]
value.} whereas Sneden \etal\ (1992\markcite{SKPL92}) 
did find variations in these elements in M5 ([Fe/H]~= --1.2) similar to 
those in more metal-poor clusters such as M13, M15, and NGC~3201.  
However, the prediction fails to explain the differences between the 
CNO abundances of NGC~288 and NGC~362 (Dickens \etal\ 
1991\markcite{DCCB91}): these two clusters have very similar 
metallicities ([Fe/H]~= --1.40 and --1.27, respectively; Zinn \& West 
1984\markcite{ZW84}), yet one shows evidence for deep-mixing 
(redistribution of C, N and O abundances) whereas the other exhibits 
little or no variation in these elements. However, a new study 
by Shetrone (1999\markcite{She99}) indicates that both clusters, in 
fact, show wide variations among the C, N, and O elements, and that the 
earlier conclusions in declaring no variations in NCG~288 resulted from 
a limited sample size.

Variations in the cyanogen band strengths have also been studied 
extensively in the giant stars of old globular clusters, as well as 
nearby dwarf spheroidal galaxies.  CN variations have been found in 
NGC~362, M5, M10, NGC~6352, NGC~7006, M92, M15, M71, M22, $\omega$~Cen, 
NGC~3201, NGC~6752, 47~Tuc and M4 (see eg. Smith \& Norris 
1982\markcite{SN82} and references therein) as well as in the Sculptor 
(Smith \& Dopita 1983\markcite{SD83}) and Draco (Smith 
1984\markcite{Smi84}) dwarf galaxies.  In clusters where a range in CN 
strengths is observed, there appears to be a correlation with elements 
such as oxygen, sodium, and aluminum.  Examples include 47~Tuc and 
NGC~6752 (Cottrell \& Da Costa 1981\markcite{CD81}) and M22 (Smith \& 
Wirth 1991\markcite{SW91}), clusters spanning a metallicity range that 
encompasses M4.  And, in some of the systems (NGC~6752, M22, 
$\omega$~Cen, and the dwarf galaxies), variations in calcium positively 
correlate with variations in cyanogen band strengths.  Norris \& 
Bessell (1978\markcite{NB78}) were among the first to recognize the 
resemblance in chemical signatures between giants in M22, $\omega$~Cen, 
and the Sculptor and Ursa Minor dwarf galaxies.  Extensive abundance 
studies of the three most interesting globular cluster systems of this 
kind have been done by Norris \etal\ (1981\markcite{NCFD81}, NGC~6752), 
Twarog \etal\ (1995\markcite{TTC95}, M22) and Norris \etal\ 
(1996\markcite{NFM96}, $\omega$~Cen).

Given the importance of decoupling the evolutionary effects from 
primordial enrichments, these puzzles led us to consider an abundance 
study of a large sample of bright giants in the mildly metal-poor 
globular cluster M4.  M4 (NGC~6121) is possibly the nearest globular 
cluster ($d$~$\sim$~1.7--2.1~kpc; see Dixon \& Longmore 1993 for a 
review).  Despite the fact that M4 lies behind the outer portion of the 
Scorpius-Ophiucus dust cloud complex and thus suffers relatively high 
reddening and extinction, its most luminous giants have relatively 
small apparent magnitudes and are readily accessible to detailed 
analysis. M4's metallicity on the Zinn \& West (1984\markcite{ZW84}) 
scale is [Fe/H]~=~--1.3.  High-resolution spectroscopic estimates 
range from [Fe/H]~= --1.3 (Brown \etal\ 1990\markcite{BWO90}; Gratton 
\etal\ 1986\markcite{GQO86}) through --1.2 (Brown \& Wallerstein 
1992\markcite{BW92}) to --1.05 (Drake \etal\ 1992\markcite{DSS92}).
Liu \& Janes (1990\markcite{LJ90}) discuss M4 reddening and metallicity 
issues at length, and Drake \etal\ (1994\markcite{DSS94}) provide a 
useful summary table of the many metallicity estimates for this cluster.
They compute a mean M4 metallicity, excluding their own result, of 
$<$[Fe/H]$>$~=~--1.17.  This metallicity is therefore near the 
anticipated cutoff in the deep-mixing process that leads to large light 
element abundance variations in red giant branch (RGB) stars of other 
clusters. 

Norris (1981\markcite{No81}) discovered CN bimodality in M4: stars of 
similar luminosity exhibit a largely bimodal distribution of cyanogen 
strengths. Norris suggested that the distribution of CN band strengths 
among cluster giants could be related to the color distribution of 
stars on the horizontal branch (HB), {\it i.e.}, the so-called 
``second parameter'' problem ({\it e.g.}  Smith \& Norris 
1993\markcite{SN93}).  It is not clear, however, whether CN bimodality 
results from deep-mixing (variations in C$\rightarrow$N conversion 
among giants) or reflects instead some built-in primordial difference 
among cluster stars. Norris argued for a range in core rotational 
velocities inducing variable amounts of envelope mixing among the 
former main sequence stars that we now see as the M4 giants.

Previous abundance studies of the brighter giants in M4 present a 
rather complex picture.  Brown \& Wallerstein (1989\markcite{BW89})
showed that C was depleted and N enhanced in four stars.  They and 
Smith \& Suntzeff (1989\markcite{SS89}) found that M4 giants all had 
very low carbon isotope ratios, nearly the equilibrium ratio of 
\carbiso~$\simeq$ 3.5.  This value is much lower than the ratio of 20 
to 30 expected from the ``first dredge-up'' mixing episodes of Pop~I 
giants (Iben 1964\markcite{Ib64}), and suggests the existence of a 
mixing mechanism more efficient than classical convection. 
Additionally, the Drake \etal\ (1992\markcite{DSS92}, 
1994\markcite{DSS94}) analysis of two pairs of CN-strong and CN-weak 
giants found that the CN-weak pair had ``normal'' low abundances of 
Na and Al ({\it i.e.}, normal in comparison to field halo giants
of comparable metallicity), whereas the CN-strong pair had 
significantly enhanced abundances of Na and Al. The CN-strong pair 
also had slightly ($\sim$~0.2~dex) lower [O/Fe] ratios than the 
CN-weak pair.  These results are compatible with the deep-mixing 
hypothesis described above (Langer \etal\ 1993\markcite{LHS93}, 
Langer \& Hoffman 1995\markcite{LH95}, Langer \etal\ 
1997\markcite{LHZ97}).

On the other hand, Brown \etal\ (1990\markcite{BWO90}) found that M4 
RGB stars have a relatively small (possibly insignificant) range of 
oxygen abundances compared with other clusters. Thus if deep-mixing 
is responsible for the changes in ${\rm ^{12}C}$, ${\rm ^{13}C}$ and 
N, it may be that the mixing is not deep enough to penetrate those 
layers of the hydrogen-burning shell in which significant burning of 
O$\rightarrow$N takes place. Similarly, Brown \& Wallerstein 
(1992\markcite{BW92}) noted that their sample of M4 giants all had 
significant overabundances of Na (see also Gratton \etal\ 1986 and 
Gratton 1987\markcite{Gra87}) and Al, but not the corresponding 
depletions of O and Mg that might be expected if mixing had been deep 
enough to bring up the products of ON, NeNa and MgAl cycling.

To summarize, M4 RGB stars have been subjected to high resolution 
spectral analysis in three independent investigations: Brown \& 
Wallerstein (1992 and references therein; seven stars); Drake \etal\ 
(1992, 1994; four stars); and Gratton \etal\ (1986; three stars).
However, there is little agreement among these investigations about 
the behavior of light element abundances.  Unfortunately, the only 
overlap among these samples is M4 L3624 (Wallerstein, Leep, \& Oke 
1987\markcite{WLO87}; Drake \etal\ 1992, 1994).  Moreover, the data 
sets and analysis techniques vary substantially among the 
investigations, so possible systematic effects in the abundances 
cannot easily be explored. The total sample size for M4 remains 
small.

We therefore have used the \'echelle spectrographs of McDonald and 
Lick Observatories to gather high resolution, large wavelength 
coverage, high signal-to-noise spectra of 25 M4 giant stars.  Our 
spectrum analysis methodology yields values of \teff\ independent of 
the photometric colors.  This provides an independent measure of
the reddening to each program star as well as the total-to-selective 
extinction ratio in the part of the dark cloud that obscures the cluster, 
a quantity believed to be anomalous ({\it e.g.} the review by Dixon \& 
Longmore 1993\markcite{DL93}).  We have derived N and O abundances and 
\carbiso\ ratios, abundances of the light elements Na, Mg, Al, Si, Ca, 
and Ti, the Fe-peak elements Sc, V, Fe and Ni, and the neutron-capture 
elements Ba and Eu.  Additionally, we have used a multi-object 
spectrometer at the Cerro Tololo Interamerican Observatory to obtain 
medium-resolution spectra of 24 M4 giants.  Thirteen of the stars in the 
medium resolution survey are also part of our high resolution sample, 
and the other eleven are generally giants of lower luminosity.  These 
spectra have limited wavelength coverage, centered near \wave{6300}, 
from which we derived [O/Fe] ratios.  With this large sample of M4 
giants we explore the distribution of several key light and heavy element 
abundances in M4.

\section{Observations, Reductions, and Equivalent Width Measurements}

M4 is possibly the nearest globular cluster ($d$~$\sim$~1.7--2.1~kpc;
see Dixon \& Longmore 1993 for a review).  Its giants are unusually 
bright; the RGB tip reaches $V$~$\sim$~10.5.  The cluster lies 
projected on rich star fields somewhat toward the Galactic nuclear 
bulge.  Fortunately, proper motion (Cudworth \& 
Rees 1990\markcite{CR90}) and radial velocity (Peterson \etal\ 
1995\markcite{PRC95}) studies have made selection of true M4 members 
fairly easy for stars as faint as two magnitudes below the HB.  
Assuming that M4 giants were only those with Cudworth \& Rees 
membership probabilities $\geq$99\%, 
we acquired high and medium resolution spectra for all 21 RGB stars 
with $V$~$<$~12.0, and a total of 36 out of the 76 stars with 
$V$~$<$~13.1.

The program stars are listed in Table~\ref{Ivans.tab1}; Lee 
(1977\markcite{Lee77}) identification numbers are used here and 
throughout this paper, with the exception of M4 G273, unobserved by 
Lee.  Alternate star names with just a letter or a prefix A are those 
of Alcaino (1975\markcite{Al75}), names with a prefix V are those of
Sawyer Hogg (1973)\markcite{SH73}, and names with a prefix G are those 
of Greenstein (1939\markcite{Gr39}).  The $V$ and $B-V$ values are 
taken from the photographic photometry of Cudworth \& Rees 
(1990\markcite{CR90}).  In Figure~\ref{Ivans.fig1} we reproduce the 
total Cudworth \& Rees M4 c-m diagram down to a $V$ magnitude of 15.4 
and show an enlargement of the region with $V$~$<$~13.1.  This figure 
demonstrates the evolutionary domain of our program stars; the points 
in the enlargement are coded by the observatory and spectrograph 
resolution used to obtain each spectrum.

\subsection{High Resolution Spectra}

We gathered the high resolution spectra in 1997 using two \'echelle 
spectrographs equipped with large-format CCD detector systems:
the McDonald Observatory 2.7m ``2d-coud\'e'' (Tull {\it et al.} 
1995\markcite{TMSL95}), and the Lick Observatory 3.0m ``Hamilton''
(Vogt 1987\markcite{Vo87}, Valenti \etal\ 1995\markcite{VBM95}).  
A standard set of tungsten filament lamp, Th-Ar hollow cathode lamp,
``bias'' and ``dark'' integrations accompanied the program star spectra. 
We also obtained spectra of several hot, rapidly rotating (essentially
featureless) stars of similar airmass to M4.  The effective wavelength 
range in both sets of spectra was 5200~{\rm \AA}~$ \leq \lambda \leq 
$~8800~{\rm \AA}.  The resolving power R $\equiv \lambda/\Delta\lambda$, 
as measured from the Th-Ar spectra, was $\simeq$60,000 for most of the 
2d-coud\'e data and $\simeq$50,000 for the Hamilton data.  A few fainter 
M4 giants were observed with the 2d-coud\'e at lower spectral resolving 
power (R~$\simeq$~30,000).  The lunar cycle was a concern; in order to 
minimize the amount of scattered moonlight in our spectra we avoided 
dates when the moon would be especially bright and close to M4 in the 
sky.  Table~\ref{Ivans.tab1} lists dates and places of the observations, 
total exposure times (which sometimes represent the sum of two or three 
individual integrations), as well as the estimated signal-to-noise 
ratios in the final co-added reduced spectra near \wave{6350}; for the 
whole sample, $<$S/N$>$~$\simeq$~80.

Initial spectrum reductions included correction for the CCD frame bias 
level, trimming of the frame overscan region, division by flat field 
frames, interpolation over bad pixels and anomalous radiation events, 
removal of scattered light underlying the spectral orders, extraction of 
the individual orders, and implantation of a wavelength scale with the 
aid of the Th-Ar integrations.  These we accomplished using standard 
routines in IRAF\footnote{IRAF is distributed by the National Optical 
Astronomy Observatories, which are operated by the Association of 
Universities for Research in Astronomy, Inc., under cooperative agreement 
with the National Science Foundation.} and, for the Hamilton data, VISTA 
(Goodrich \& Veilleux 1988\markcite{GV88}) software packages; see previous 
papers in this series (\eg, Sneden \etal\ 1991\markcite{SKPL91}, Kraft 
\etal\ 1993\markcite{KSLS93}) for further discussion of these procedures.

The resulting singly-dimensioned spectra were processed further with 
specialized software (SPECTRE: Fitzpatrick \& Sneden 1987\markcite{FS87}) 
in order to compute radial velocities, continuum-normalize the spectra, 
remove telluric spectral features, and measure equivalent widths.  During 
these tasks it became apparent that star L1412 could not be treated in 
the same manner as the other program stars.  Our spectra reveal much 
broader absorption lines in L1412 than in any other M4 giant and one that 
varied significantly over a one-year period, not only in radial velocity 
but in line depth ratios of temperature-sensitive lines.  At $V$~=~10.38, 
this star is by far the brightest M4 cluster member, and the aberrant 
nature of its spectrum (particularly the very strong CN and detectable 
TiO bands) has been well documented in the literature.  Whitmer \etal\ 
(1995\markcite{WBBW95}, and references therein) have recently discussed 
L1412 in some detail, concluding that it is a post-AGB star.  This star 
would have required a very different analysis than that employed for the 
rest of our sample, and so we dropped it from further consideration in 
this study.

We used the stellar radial velocities v$_{\rm r}$ to recheck cluster 
membership for our program stars and to compare with velocity estimates 
in the literature.\footnote{Our preliminary reconnaissance of the 
spectrum of M4 L4513, suggested to be a 99\% 
probable member from the star's proper motion (Cudworth \& Rees 1990),
revealed a discordant radial velocity (more than 110~\kmsec\ blueshifted 
with respect to the cluster mean velocity).  Peterson \etal\ (1995) also 
noted this star's discrepant velocity, rejecting it as a cluster member.
Additionally, the \teff\ we derived for this star (with an identical 
analysis to the other M4 targets) produced implied colors and absolute 
magnitudes very different than those of the rest of our sample.  We 
therefore discarded L4513 from this study.}  The v$_{\rm r}$ values were 
computed by comparing the measured line-center wavelengths of 13 lines in 
the \wave{6300} order to those listed in the catalogue of solar lines by 
Moore \etal\ (1966)\markcite{MMH66} and converting to velocity units.  
Similar measurements of telluric O$_2$ line wavelengths on the same 
\'echelle order set the zero point of the velocity scale for each spectrum.
Each line center was determined by eye to about 1/3 pixel 
($\sim$~2~\kmsec) accuracy.  For stars that were observed twice in one 
night, our mean v$_{\rm r}$ determinations agree very well, with a 
scatter of 0.1~\kmsec.  The mean v$_{\rm r}$ and the standard deviation 
for each star are entered in Table~\ref{Ivans.tab1}, along with the values 
of Peterson \etal\ (1995\markcite{PRC95}).  The agreement with their work 
is excellent: taking the differences in the sense {\it this study minus 
Peterson et al.}, $<\delta$v$_r>$~= --0.18~$\pm$~0.23~\kmsec\ ($\sigma$~= 
0.99~\kmsec, for 21 stars in common).

M4, with a declination of --26.5$^o$, never rises higher than an
altitude of $\simeq$32$^o$ at McDonald Observatory and $\simeq$25$^o$ 
at Lick Observatory.  Telluric O$_2$ and H$_2$O spectral lines often are 
very strong at such large airmasses, and special care must be taken in 
the cancelation of these features via division by the ``telluric standard''
hot-star spectra.  With a careful choice of divisor star (very early 
spectral type, large $vsini$, and best airmass match to M4), and through 
modest manipulation of the divisor telluric line strengths in SPECTRE, we 
achieved essentially complete removal of telluric features.  We illustrate 
this by one example shown in Figure~\ref{Ivans.fig2}.  But, even with such a 
robust procedure, residual telluric feature contamination might be of 
concern for the analysis of the important \wave{6300.3} [\nion{O}{i}] 
line.  Fortunately, the average cluster radial velocity of 
$\sim$~+70~\kmsec\ translates to a red shift of more than 1~{\rm \AA}, 
leaving the stellar [\nion{O}{i}] well clear of any O$_2$ absorption 
line, as well as the night sky [\nion{O}{i}] emission line.

Even though M4 has a metallicity of less than one-tenth solar, its 
upper RGB stars have strong and complex spectra; special care must be 
taken in continuum placement.  For each spectral order the continuum was 
set by interactively fitting a spline function to line-free spectral 
regions.  Locating the continuum was aided by comparing the spectra of 
our program stars with that of Arcturus (Griffin 1968\markcite{Gr68}).

We initially adopted the atomic line lists used in previous papers of
this series, but after careful line-by-line inspection of both our 
spectra and that of the Arcturus atlas, we culled four lines which seem 
to be too blended to be used in cool RGB stars having M4's metallicity.
Unfortunately we were then left with a list of only fairly strong 
lines, and so we added as many clean lines of low--medium strength as 
were available from Langer \etal\ (1998\markcite{LFSB98}), Lambert 
\etal\ (1996\markcite{LHLD}), and Th\'evenin (1990\markcite{Th90}).  We
also included in our line lists the \nion{La}{ii} lines of Brown \&
Wallerstein (1992\markcite{BW92}) and these line choices will be 
discussed further in \S3.5.  We measured equivalent widths (EWs) for 
these lines with SPECTRE, using Gaussian approximations for all but the 
strongest lines, for which Voigt profile fits were employed.  The tables 
of EWs and associated atomic parameters are too large to include here, 
but are available electronically from the first author.   We did not 
analyze any line with ${\rm log}{\rm (EW/\lambda)}>$ --4.25 with the 
exception of those of \nion{Ba}{ii}.  Such strong lines are too 
sensitive to choices in both microturbulent velocity and line damping 
constant to yield reliable abundances.  We did retain lines of 
\nion{Ba}{ii} (which has only very strong transitions); those were 
treated with synthetic spectrum computations as described in \S3.

Our EWs agree reasonably well with those in the literature.  Taking 
comparisons in the sense {\it this work minus others}, for the four 
stars in common with Drake \etal\ (1992), $<\delta$EW$>$~= 
+1.4~$\pm$~0.8~m{\rm \AA}\ ($\sigma$~=~5.2~m{\rm \AA}, 43 lines).
We also made independent measurements of some of the lines used in the 
Drake \etal\ study, as well as some of the additional lines used in 
the present study.  Comparison with those measures revealed 
consistency with ours: $<\delta$EW$>$~= +2.5~$\pm$~0.7~m{\rm \AA}\ 
($\sigma$~=~4.4~m{\rm \AA}, 45 lines).  We have three stars in common 
with Brown \etal\ (1990) neglecting L1412, and Brown 
(1998\markcite{Br98}) has kindly made available to us their 
unpublished EW measurements.  Our EW scale has a small negative 
offset from theirs, and there is significant line-to-line scatter: 
$<\delta$EW$>$~= --6.9~$\pm$~1.3~m{\rm \AA}\ 
($\sigma$~=~10.9~m{\rm \AA}, 66 lines).  We suspect that the offset 
arises from the large difference in spectral resolving powers of the 
two studies: our spectra of these stars were obtained at 
R~$\simeq$~60,000, while the Brown \etal\ spectra had 
R~$\simeq$~20,000.  There are CN lines throughout the red spectral 
region, and it is difficult to avoid inclusion of one of them in EW 
measurements.  Our higher spectral resolution undoubtedly aided in 
spotting CN lines lurking in the wings of some atomic lines, hence
our mean EWs would be expected to be lower.

\subsection{Medium Resolution Spectra}

In 1992 we surveyed M4's RGB with the CTIO 4m Argus fiber-coupled 
multi-object spectrometer (Ingerson 1993\markcite{In93}).  The 
spectra were obtained to study the strongest [\nion{O}{i}] feature, 
and so were centered at \wave{6300}.  The spectral coverage was 
approximately 6240~$< \lambda <$~6360~{\rm \AA}, the reciprocal 
dispersion was $\simeq$~0.2~{\rm \AA}/pixel, and the effective 
resolving power was R~$\simeq$~20,000.  We obtained spectra of 24 
M4 giants and used the 24 remaining Argus fibers to gather sky 
spectra.  As with the high resolution data, we acquired companion 
spectra of a flat-field lamp source, a hot rapidly-rotating star, 
and a Th-Ar emission line lamp.  The flat-field lamps were projected 
onto the 4-m dome and observed through the telescope light path.  We 
also observed the twilight sky with Argus to aid in the wavelength 
identifications.

The Argus frames were reduced to one-dimensional 
wavelength-calibrated spectra with standard IRAF routines (see 
Suntzeff \etal\ 1993\markcite{SMTOGW93} for a detailed discussion of 
the reduction of the Argus fiber data).  The data reduction included 
sky subtraction with mean spectra chosen from the 24 sky fibers. 
Argus has a substantial focus change going from fiber 1 to 48 along 
the entrance slit to the spectrograph (perpendicular to the dispersion), 
resulting in a detectable spectral resolution difference among the 
fibers.  The sky spectra were formed from the seven sky fiber spectra 
nearest to the program fiber spectra.  The M4 data were observed in 
three exposures. The reduced data were co-added and the final spectra 
represent a total integration time of 95 minutes.

Radial velocities were measured with respect to a single velocity 
template.  All the fibers were put on a common relative system for a 
single CCD exposure, and then all of the frames during the night were 
put on the same relative system by cross-correlating all the Th-Ar 
spectra and applying a zero point to each frame.  We then demanded that 
the average velocity of M4 be the same as the average velocity of the 
Peterson {\it et al.} (1995) stars observed in common, bringing our 
velocities onto their absolute velocity system.  The observed 
dispersion about the (defined zero) mean was 0.8~\kmsec\ with respect 
to Peterson {\it et al.}

We used SPECTRE to carry out telluric line cancelation.  The telluric
standard hot-star spectra were obtained in multiple fibers.  For each 
program star spectrum, we co-added (to increase S/N) the reduced telluric
hot-star spectra from several of the nearest fibers surrounding the slit 
position of the program star fiber before applying the telluric 
cancelation routines.

In Figure~\ref{Ivans.fig3} we show two fully reduced Argus 
medium-resolution spectra of stars for which we also obtained high 
resolution data.  The stars chosen for display illustrate the 
appearance of the spectra (intrinsic line strengths and S/N of the 
data) for a star near the RGB tip and one near the low-luminosity limit 
of the high-resolution targets.  The [\nion{O}{i}] \wave{6300.3} line is 
marked, and we also draw attention to several \nion{V}{i} and 
\nion{Ti}{i} lines in this spectral region.  Their large sensitivity to 
$B-V$ (hence \teff) is apparent from these spectra.  However, none of 
these lines are among those recommended by Gray (1994) for line-depth 
ratio studies (potential contamination by O$_2$ telluric lines limit 
their utility for ultra-precise \teff\ measurements).   The sacrifice of 
resolution in order to gain stellar sample size is easily seen in the 
Argus spectra of this figure.  There are few if any truly unblended 
spectra features over the 120~{\rm \AA}\ spectral range of these data.
Therefore we measured no EWs on the Argus spectra and performed no
line-by-line abundance analysis.  However, the [\nion{O}{i}] line is 
clearly detected in all Argus targets, and we were able to derive 
reliable [O/Fe] ratios from these data via synthetic spectrum analyses, 
after first analyzing the high resolution data and using those results 
to anchor the computations for the medium resolution spectra.

\section{Abundance Analysis}

In this section we describe the derivation of abundances in our samples 
of M4 giants.  We consider two very different spectroscopic data sets 
for a cluster that presents unusual analytical difficulties irrespective 
of the data characteristics.  Analysis of the high resolution data is 
discussed in the first several subsections, and the last subsection is 
devoted to the medium resolution data.

\subsection{The Special Challenges of M4 Giant Star Spectra}

The translation of EWs from high resolution spectra into reliable 
abundances depends on the accurate determination of four stellar 
atmosphere parameters: effective temperature \teff, gravitational 
acceleration \logg, overall metallicity [Fe/H], and microturbulent 
velocity \vmicro.  A distinct advantage of most globular cluster 
abundance analyses is the ability to use cluster c-m diagram photometry 
to specify preliminary values of \teff\ and \logg, thereby leaving 
essentially only [Fe/H] and \vmicro\ as free parameters to be determined 
from spectral line analysis.  We have used this technique in many of our 
previous papers, but it cannot be done in M4 (nor can it be employed for 
many higher metallicity ``disk population'' globular clusters).  The 
insurmountable difficulty here lies in the large and differential 
reddening across the face of M4 produced by the complex Sco-Oph dust 
distribution.  Reddening cannot reliably be estimated for individual M4 
stars to the level needed to map broad-band photometric indices onto 
stellar parameters.  The spatial variability of reddening in M4 does not 
permit even a reliable photometric {\it ranking} of our stars in order 
of \teff.  Moreover, all bright M4 giants have strong-lined spectra, and 
that adds to the frustrations:  cleanly measurable absorption lines 
usually lie on the flat (saturated) part of the curve-of-growth, so that 
plausible abundances may be obtained for a variety of assumed \teff, 
\logg\ and \vmicro\ combinations.  A different strategy must be employed, 
and after conducting a number of numerical experiments we constrained M4 
stellar effective temperatures with the following technique.

We first adopted a modified version of the method originally described 
by Gray \& Johanson (1991\markcite{GJ91}) that uses the ratios of 
central depths of lines having very different functional dependences
on \teff\ to derive accurate relative temperature estimates.  The \teff\ 
calibration of the M4 line depth ratios was set through a similar 
analysis of RGB stars of M5 (a cluster of very similar metallicity to M4 
but little reddening problem).  Then, in order to anchor the \teff\ scale 
and to lift most of the remaining stellar parameter degeneracies, we 
demanded that abundances determined from \nion{Ti}{i} and \nion{V}{i} 
lines yield [Ti/Fe] and [V/Fe] abundance ratios that (a) did not 
appreciably vary from star-to-star in M4, and (b) were roughly consistent
with those ratios that have been derived in our previous studies of other, 
more easily analyzed globular clusters.  \nion{Ti}{i} and \nion{V}{i} have 
much lower ionization potentials than does \nion{Fe}{i}, and available 
spectral features of these two species arise from lower excitation states 
than do the typical \nion{Fe}{i} lines we employed.  Their derived 
relative abundances are thus very sensitive to \teff.  We further demanded 
that the [Fe/H] abundance ratios fall within the range 
--1.05~$\geq$~[Fe/H]~$\geq$~--1.30, as suggested by the earlier M4 
analyses cited in \S1.  However, [Ni/Fe] was left free to vary from any 
and all constraints.  Finally, we estimated gravities by obtaining 
equality in abundances of neutral and ionized species of both Fe and Ti.

The ``Gray'' method is a purely {\it empirical} match between measured 
line depth ratios and photometric indices and/or \teff\ scales; its use 
requires no defense here.  More justification is needed for our other 
assumptions in stellar model determinations.  If the chief goal of this 
work is to search for star-to-star element abundance variations, how can 
we insist on little scatter in the relative Ti and V abundances?  The 
arguments here are both theoretical and observational.  First, V and Ni 
are part of the ``Fe-peak'', and are synthesized together with Fe during 
the most advanced quiescent and explosive core fusion stages of high mass 
stars.  Ti has a more complex nucleosynthetic origin, being produced both 
as an Fe-peak element and as the heaviest identifiable ``$\alpha$-capture'' 
element.  In field metal-poor stars, at least over the metallicity range 
--1.0~$\geq$~[Fe/H]~$\geq$~--2.5, observers agree that 
[V/Fe]~$\simeq$~[Ni/Fe]~$\simeq$~0.0, while [Ti/Fe]~$\simeq$~+0.3 (\eg, 
Gratton \& Sneden 1991\markcite{GS91}, and references therein).  The 
star-to-star scatters in these means are very small, consistent with 
being dominated by observational errors.  Corresponding cluster 
abundances are in complete accord with the field star values.  From 
earlier papers in this series, straight means of the average abundance 
ratios in M3 and M10 (Kraft \etal\ 1995\markcite{KSLSB95}), M13 (Kraft 
\etal\ 1997\markcite{KSSSLP97}), M5 (Sneden \etal\ 
1992\markcite{SKPL92}), M92 (Sneden \etal\ 1991\markcite{SKPL91}), M15 
(Sneden \etal\ 1997\markcite{SKSSLP97}), and NGC~7006 (Kraft \etal\ 
1998\markcite{KSSSF98}), are: 
$<$[Ti/Fe]$>$~=~+0.26 ($\sigma$~=~0.04; no value for M15 and M92);
$<$[V/Fe]$>$~=~--0.01 ($\sigma$~=~0.04); and 
$<$[Ni/Fe]$>$~=~--0.01 ($\sigma$~=~0.09; no values for M15 and M92).
We expect these values to approximately hold in M4.

\subsection{Line Depth Ratios Applied to M4 Effective Temperatures}

Gray \& Johanson (1991) demonstrated that the ratio of the central depth 
of \nion{V}{i} \wave{6251.83} to that of \nion{Fe}{i} \wave{6252.57} is 
correlated very strongly with $B-V$ and \teff\ in Pop~I main sequence 
F7--K7 stars.  Their work was developed by Gray (1994\markcite{Gr94}) to 
include more line pairs, and Hiltgen (1996\markcite{Hi96}) further 
extended the method to G--K subgiants over a larger range of disk 
metallicities.  Happily, many of Gray's line depth ratios are also 
sensitive \teff\ indicators for lower metallicity very cool RGB stars.
We illustrate this in Figure~\ref{Ivans.fig4} with reproductions of 
spectra of two M4 stars that have $\delta\teff~\sim $~700~K.  In this 
temperature regime, the strengths of \nion{V}{i} lines clearly increase 
with decreasing \teff\ while \nion{Fe}{i} lines change very little in M4 
RGB stars.  Other features employed by Gray change more slowly, but all 
ratios retain some sensitivity to \teff.

In Figure~\ref{Ivans.fig5} we show the correlations of two Gray 
ratios with temperature in three clusters previously studied in this 
series.\footnote{A comprehensive investigation of the spectroscopic line
ratios of these data sets is underway and will be reported in a future 
publication.}  The illustrated ratios are d8/d7 (or \nion{V}{i} 
\wave{6223.20}/\nion{Fe}{i} \wave{6229.23}) and d5/d6 (or \nion{V}{i} 
\wave{6224.51}/\nion{Fe}{i} \wave{6226.74}).  The \teff\ values are 
taken from Sneden \etal\ (1992)\markcite{SKPL92} for M5, Kraft \etal\ 
(1995)\markcite{KSLSB95} for M10, and Kraft \etal\ 
(1997)\markcite{KSSSLP97} for M13.  Data for M4 also are included in 
this figure, and they are developed in the present and next section.
Figure~\ref{Ivans.fig5} shows quantitatively how the (base-10 
logarithm of) line depth ratios vary more than one dex in spectra of 
giants of moderately metal-poor clusters, and thus can indicate very 
small \teff\ changes.  These relationships begin to flatten out among the 
coolest stars, as both dividend \nion{V}{i} and divisor \nion{Fe}{i} line 
strengths begin saturating, and their central depth ratios approach unity.  
Thus these ratios probably will be less useful as temperature indicators 
for the coolest stars of appreciably more metal-rich globular clusters.  
Clusters with [Fe/H]~$>$~--1, such as 47~Tuc and M71, have redder (thus 
cooler) RGBs than M5 and M4, and the combination of lower \teff\ and 
higher [Fe/H] conspire to saturate and blend virtually all of the usual 
spectral features.

Turning the line depth ratios into temperatures requires some 
calibration steps, and here we describe the procedure used for M4.  We 
first considered M5, a cluster whose metallicity ([Fe/H]~=~--1.17; 
Sneden \etal\ 1992) is similar to M4, but which suffers very little
interstellar dust extinction ($E(B-V)$~=~0.03, or $A_V \simeq$~0.1;
Frogel \etal\ 1983).  We measured all line depth ratios for M5 
recommended by Gray (1994) that appear on the Hamilton spectra of 13 
stars previously treated by Sneden \etal\ and 20 stars newly observed 
with the Keck ``HIRES'' \'echelle spectrograph.\footnote{A comprehensive 
analysis of these M5 data sets will also be reported in a future 
publication.}  We correlated the line depth ratios against de-reddened 
M5 $B-V$ values from Cudworth (1979\markcite{Cud79}) and Simoda \& 
Tanikawa (1970\markcite{ST70}), and fit linear or quadratic regression 
lines to these data to create formulae to predict $B-V$'s from the depth 
ratios.

We then measured the line depth ratios in our M4 stars and used the
M5 formulae to predict a \BmV0\ value for each ratio in each M4 star.
Weighted means were then computed for each star, with the weights
directly proportional to the slopes of the logarithmic depth ratio 
{\it vs.} $B-V$ regression relationships and inversely proportional to 
the observed scatters around these relationships.  These implied 
colors were compared to the M5 temperatures derived by Sneden \etal\ 
(1992) to produce initial M4 \teff\ estimates.  Such \teff's were 
expected to be accurate for {\it ranking} the stars; zero-point (or 
slope) offsets from M5 \teff's remained to be determined from full 
spectral analyses.

\subsection{Final Model Parameters}

As discussed in \S3.1, we used line-by-line abundance analyses to 
complete the model atmosphere specifications.  We employed the current 
version of the LTE line analysis code MOOG (Sneden 1973\markcite{Sn73}) 
for these computations.  As in earlier papers of this series, we input 
to MOOG trial model atmospheres generated with the MARCS code 
(Gustafsson \etal\ 1975\markcite{GBEN75}) and the atomic line data 
described in \S2.  A few comments on the line lists should be made 
here.  First, Ni and Fe have similar first ionization potentials and 
atomic structures.  We expected the scatter in [Ni/Fe] to provide 
additional information regarding how well the [Fe/H] abundances had 
been determined.  Therefore a special effort was made to identify as 
many clean \nion{Ni}{i} lines as possible; we eventually used nine 
lines in most stars.  Second, many \nion{V}{i} transitions have large 
hyperfine splitting.  Since these lines often are very strong in M4 
giants, we limited our analysis to \nion{V}{i} $\lambda\lambda$6274.6 
and 6285.1~{\rm \AA}, two lines with well-determined laboratory hyperfine 
structure components (as reported by McWilliam \& Rich 
(1994\markcite{MR94})), normalizing the gf-values to those adopted for 
these lines in our previous work.  We used the blended-line option of 
MOOG to derive abundances from the EWs of these lines.  Finally, we 
added the \wave{6606.9} line of \nion{Ti}{ii} to our previously 
employed \nion{Ti}{i} lines, in order to provide a second ionization 
equilibrium check.

Iterative abundance runs with small excursions in \teff, \logg, model 
[M/H], and \vmicro\ yielded the final model parameters; these are 
listed in Table~\ref{Ivans.tab2}.  Recalling from \S3.1 that there 
is much coupling among the model parameters and no assistance to be 
had from photometry, we concentrated on the following abundance 
indicators for the model iterations.  For \teff: no obvious trends of 
\nion{Fe}{i} abundances with excitation potential; sensible Ti and V 
abundances ([Ti/Fe]~$\sim$~+0.3, [V/Fe]~$\sim$~0.0); and Fe abundances 
within the range --1.05~$\geq$~[Fe/H]~$\geq$~--1.30 (again, as 
suggested by several previous analyses).  For \logg: consistent 
abundances from lines of neutral and ionized species of Fe and Ti; 
reasonable predicted c-m diagram positions with the derived gravities 
(see \S 3.4).  For \vmicro: no obvious trends of \nion{Fe}{i} line 
abundances with EWs.  The whole set of iterated models was rechecked 
for consistency with the \teff\ rankings derived from the initial
line depth ratio calibrations.  Overall, we demanded that the Ti, V, 
and Fe abundances show no significant drifts along the M4 RGB.

These model constraints appear to be so restrictive that they 
virtually guarantee particular sets of model parameters.  But, in 
reality, the nearly 1000~K range in \teff\ is quite helpful, for there 
is no astrophysical reason for the Fe-peak abundances to vary 
significantly from star to star along the M4 giant branch.  The model 
parameter couplings are much less severe in the warmer stars of our 
high resolution sample, and those stars yield {\it without constraint} 
the abundance ratios suggested above.  Note also that for all stars, 
the abundances deduced from \nion{Ti}{i} and \nion{V}{i} work together: 
the [Ti/V] ratios are nearly insensitive to model parameter choices.
Finally, inspection of Figure~\ref{Ivans.fig5}'s correlation
between the line depth ratios and our final \teff\ estimates for M4 
giants suggests reasonable agreement.  Our M4 \teff\ scale is on the 
same general system as that employed in our earlier investigations of 
M5, M10, and M13.

In the three coolest M4 stars (L1514, L4613, and L4611) we were unable 
to satisfy all of the constraints equally well.  The abundances for 
these stars, where they appear in the figures that follow, have been 
marked with a hollow square symbol ($\Box$).  The MARCS models we used 
for all stars assume LTE.  For these cool stars at the tip of the giant 
branch, with their extended envelopes and extremely low \logg\ values, 
formation of \nion{Fe}{i} and other lines in the atmospheres may be 
affected by departures from LTE.  If Fe and Ti are over-ionized in the 
outer layers of these very low density giants, the gravities we obtain 
by demanding ionization equilibrium will be too low.  In the abundance 
results based on the fundamental parameters we adopted for these three 
stars, the ionization equilibrium between \nion{Ti}{i} and \nion{Ti}{ii} 
and between \nion{Fe}{i} and \nion{Fe}{ii} were not satisfied equally 
well.  The abundances of V and Ti also indicate that the stellar models 
adopted for these cool stars may not be the most suitable.  It is 
possible that we reached beyond the useful limits, in temperature and 
gravity, of the MARCS model atmospheres and/or the LTE line formation 
assumptions of the analysis code for these coolest stars.

We performed numerical experiments to determine systematic and random 
errors of our model parameters.  Based on the ``degeneracy'' of the 
models to fit the data when the additional abundance constraints were not 
included in the derivation of the fundamental parameters (see \S3.1), our 
estimate of potential systematic errors are 
$\Delta$\teff/$\Delta$\logg/$\Delta$\vmicro\ 
= $125$K/$0.25$dex/$0.25$km$\cdot$s$^{-1}$.  
Imposing the additional abundance constraints, random errors which would 
still permit a stellar model to satisfy the conditions discussed in 
\S3.1 are 
$\Delta$\teff/$\Delta$\logg/$\Delta$\vmicro\ 
= $50$K/$0.15$dex/$0.15$km$\cdot$s$^{-1}$.  The additional abundance 
constraints are most sensitive to changes in \teff\ (and less sensitive
to changes in \logg\ and \vmicro) thus, the random errors in \teff\ are 
better constrained than the errors in \logg\ and \vmicro.  We also 
investigated the differences that using Kurucz (1993\markcite{Kur93}) 
model atmospheres made on our results.  These models, in general, 
satisfied the constraints with slightly lower temperatures (50K at most), 
lower \logg\ values (0.1 to 0.2~dex), and lower microturbulent velocities 
($\sim$~0.1 km/s), all values within the margin of error we determined 
for our fundamental parameters.

\subsection{Implications of the Model Atmosphere Parameters}

As an external check on the the (\teff, \logg) pairs determined in the 
line analyses we compared the adopted gravity of an M4 giant star against 
that expected from its ``evolutionary'' position in the c-m diagram.  The 
predicted evolutionary gravity was derived by combining the gravitation 
law with Stefan's law, \logg$_{evol}$ = --10.564 + 
0.4(V~-~5${\rm log}~{\rm d}$~-~A$_{V}$~+~BC) + 4${\rm log}~$\teff, where 
we have assumed a stellar turn-off and giant branch mass of 0.85M$_{\sun}$, 
a typical value for the main sequence turn-off mass of a globular cluster 
star (eg. Smith \& Norris 1993\markcite{SN93}, Kraft 1994\markcite{Kr94}).  
The symbols have their usual meanings and several standard solar quantities 
(\logg~=~4.44, \teff~=~5772, and $M_{bol}$~=~4.72) are subsumed in the 
constant.   We calculated evolutionary gravities for a range of recently 
published M4 distances and extinctions (see Dixon \& Longmore 1993 for a 
review of these values): distances 1.7~$\leq d \leq$~2.1~kpc; mean 
reddening $E(B-V)$~=~0.37; and ratios of total to selective extinction 
[$R \equiv A_{V}/E(B-V) \simeq E(V-K)/E(B-V)$] in the range of 3.1~$\leq R 
\leq$~4.0.  The latter two parameters combine to yield a total visual 
extinction in the range of 1.18~$\leq A_{V} \leq$~1.48.  We interpolated 
Worthey's bolometric corrections (1994\markcite{Wo94}) as a function of our 
adopted \teff, \logg, and metallicity.  We also adopted the formula derived 
by Cudworth and Rees (1990\markcite{CR90}) to correct for the reddening 
gradient as a function of coordinate position across the face of the 
cluster.

In Table~\ref{Ivans.tab2} we list the evolutionary gravities for each M4 
star, assuming that $d$~=2.1~kpc and $A_V$~=~1.48.  This provided the best 
overall match between the two gravity sets.  If we define 
$\delta$(\logg)~$\equiv$~\logg$_{spec}$~--~\logg$_{evol}$, 
then for this distance/extinction choice 
$\delta$(\logg)~= +0.02~$\pm$~0.01 ($\sigma$~=~0.11).  
However, other choices of these quantities also yield reasonable agreement.  
For $d$~=~2.1~kpc and $A_V$~=~1.19, $\delta$(\logg)~=~--0.16; for 
$d$~=~1.7~kpc and $A_V$~=~1.48, $\delta$(\logg)~=~--0.09; and for 
$d$~=~1.7~kpc and $A_V$~=~1.19, $\delta$(\logg)~=~--0.28.  Only with the 
assumptions of the closest M4 distance and least interstellar extinction 
do the spectroscopic and evolutionary gravities clash badly.  Note that a 
study by Peterson \etal\ (1995) based entirely on kinematical 
considerations (proper motions and radial velocities) did indeed yield a 
``short'' distance of 1.72~$\pm$~0.04~kpc.  However, any mass loss along 
the RGB would improve the agreement between the gravity estimate based on 
spectroscopy and that based on kinematical considerations.

Stellar mass loss tracers and/or chromospheric activity indicators can 
sometimes be observed in the form of core shifts in the Na~D lines as well 
as from circumstellar material in the form of emission in the cores of 
the \nion{Mg}{ii} ($\lambda\lambda$2795~{\rm \AA}) and \nion{Ca}{ii} 
$K-$lines, and in the wings of H-$\alpha$ (see \eg, Dupree \etal\ 
1992\markcite{DSL92} and references therein).  A search of the literature 
did not uncover any spectroscopic observations of the ultraviolet region of 
the M4 giants, and unfortunately, the Na~D lines on our spectra are affected 
by telluric emission; a meaningful core shift cannot be measured.  However, 
some of our stars were observed using a wavelength setting that did include 
the H-$\alpha$ region.  We inspected these spectra for H-$\alpha$ emission 
above the level of the continuum.  The results of this exercise are given in 
Table~\ref{Ivans.tab3}, along with the emission indications reported by 
other observers (Cacciari \& Freeman 1983\markcite{CF83}, Brown \etal\ 
1990, Kemp \& Bates 1995\markcite{KB95}), for stars in common with our 
sample.  We note whether the emission is observed to the red or to the 
blue or on both sides of the H-$\alpha$ absorption features.  A colon 
denotes a marginal or weak emission detection.  One star, L4611, shows 
evidence of evolution in the two decades over which its H-$\alpha$ 
emission behavior has been observed.  But, most other M4 giants show 
little or no evidence for mass loss as indicated by the H-$\alpha$ 
profiles.

Our model parameters can be used to check ISM extinction properties toward 
M4.  From the \teff\ values, we estimated intrinsic M4 $B-V$ colors by
interpolating the model atmosphere predictions of Cohen \etal\ 
(1978\markcite{CFP78}, their Table~3 and Figure~4).  By subtracting these 
intrinsic $B-V$ colors from the observed colors given in 
Table~\ref{Ivans.tab1}, we determined the $E(B-V)$ values that are listed 
in Table~\ref{Ivans.tab4}.  Excluding the post-AGB star L1412, we were 
able to estimate $E(B-V)$ in this manner for all but the two coolest M4 
stars observed at high resolution (the Cohen \etal\markcite{CFP78} model 
predictions could not easily be extrapolated to the very cool stars L4613 
and L4611).  Figure~\ref{Ivans.fig6} shows an angular distribution map of the 
positions of our M4 program stars on the sky using the $\Delta\alpha$ and 
$\Delta\delta$ coordinates (as measured from the center of the cluster) 
from Cudworth \& Rees, indicating both Lee's stellar identifications and 
our corresponding reddening determinations.  The most heavily reddened M4 
giants from our analysis lie generally in the western half of the cluster, 
in good agreement with the differential reddening results of Cudworth \& 
Rees (1990).  Our $E(B-V)$ estimates also correlate with IRAS 100 micron 
fluxes which have been corrected for zodiacal dust ontamination.

Lyons \etal\ (1995\markcite{LBKD95}) also studied the differential 
reddening across the face of the cluster using interstellar K~{\sc i} 
column density measurements towards 16 stars, all of which we observed at 
high resolution.  The Lyons \etal\ $E(B-V)$ values are listed in 
Table~\ref{Ivans.tab4}.  Both the mean value and the range of our 
$E(B-V)$ estimates are in excellent agreement with theirs.  Taking the 
differences in the sense of {\it this study minus Lyons et al.}, 
$<\delta E(B-V)>$~= +0.01~$\pm$~0.01~magnitudes 
($\sigma$~= 0.05~magnitude), for 13 stars in common.  Another study, by 
Caputo \etal\ (1985\markcite{CCQS85}) also derived mean values of 
$E(B-V)$~= 0.32--0.33, utilizing an independent method based on the 
properties of RR Lyrae variables in M4.

We also determined the reddening to the program stars using the dust 
maps made by Schlegel \etal\ (1998\markcite{SFD98}), who reprocessed the 
high resolution IRAS/ISSA 100 micron maps using the precise COBE/DIRBE 
calibrations.  Surprisingly, the reddening estimates provided from their 
maps is significantly higher than ours.  The difference, for the 22 stars 
listed in table~\ref{Ivans.tab4}, in the sense of {\it this study minus 
Schlegel et al.}, is $<$$\delta$E(B--V)$>$~= --0.17~$\pm$~0.01~magnitudes 
($\sigma$~= 0.04~magnitude).  This difference cannot wholly be attributed 
to the assumption of $R$~=~3.1 by Schlegel {\it et al}\markcite{SFD98}. 
Arce \& Goodman (1999a\markcite{AG99a}) obtained similar overestimates 
using the Schlegel \etal\markcite{SFD98} maps in their extinction study 
of the Taurus dark cloud complex (1999b\markcite{AG99b}).  In their 
cautionary note, Arce \& Goodman\markcite{AG99a} attribute the cause of 
the reddening (dust opacity) estimate discrepancy to the correlation of 
intrinsic $E(B-V)$ and Mg$_{2}$ assumed by Schlegel \etal\  (90\% 
of the elliptical galaxies used for the regression have low reddenings 
and the fit is less than good for the few galaxies with $E(B-V)~>$~0.15
mag), producing an overall overestimate of reddening values by up to a 
factor of 1.5 in regions of large extinction. These overestimates
agree with those we found for our M4 program stars.  In another recent
study, Woudt (1998\markcite{Wou98}) compares reddening estimates using 
the Dn-Sigma relation of galaxies in an Abell cluster (ACO 3627) at low 
galactic latitude with those provided by the Schlegel \etal\markcite{SFD98} 
dust maps.  In the galaxy studied closest to the galactic plane (WKK5345), 
Woudt\markcite{Wou98} also finds that the Schlegel \etal\markcite{SFD98} 
dust maps overestimate the reddening.  However, for $E(B-V)$ values of 0.3, 
values comparable to M4, Woudt\markcite{Wou98} did not find any 
distinguishable difference between the two estimates.  Until such time as 
the cause of the possible overestimation is resolved, we echo the warning 
issued by Arce \& Goodman\markcite{AG99a} to use caution when using the 
Schlegel \etal\markcite{SFD98} dust maps to estimate reddening and 
extinction in regions of high extinction.

Because of the methods employed to derive the spectroscopic estimates of
reddening, we are able to make star-by-star comparisons of our results 
against the work done by Lyons \etal\ (1995).  For L2206, L2307, and L2406 
our reddening estimates are significantly higher than those derived by 
Lyons {\it et al}.  These three stars are in a region on the sky for which 
higher than average IRAS 100~\micron\ flux values have also been measured. 
The stars we observed nearest in the sky to these (\eg, L2208) also have 
relatively high reddening values.  It may simply be that the gas measured 
in the K~{\sc i} column density measurements---the basis of the Lyons 
\etal\ reddening estimates---does not completely trace the dust or the 
IRAS flux measurements, due to shielding effects and variations in the 
optical depth of the gas as suggested by de~Geus \& Burton 
(1991\markcite{DGB91}) in their study of the Sco-Oph molecular cloud 
region.

In addition to large and differential reddening, there is evidence that
the dust along the line of sight to M4 has anomalous absorption 
properties that deviate from the ``normal'' interstellar extinction law 
($R$~$\sim$~3.1; Harris 1973\markcite{Har73}, Schultz \& Wiemar 
1975\markcite{SW75}, Barlow \& Cohen 1977\markcite{BC77}, Sneden \etal\ 
1978\markcite{SNGHYS78}).  Cudworth \& Rees (1991), using their 
photometric observations (by minimizing the spread in $V$ in the HB), 
found $R$~= 3.3~$\pm$~0.7.  Work done by Clayton \& Cardelli 
(1988\markcite{CC88}) gives $R$~=~3.8 for $\sigma$ Sco, a star only one 
degree in the plane of 	the sky away from M4.  Studying the outer parts 
of the Sco-Oph dust cloud complex, Chini (1981\markcite{Ch81}) proposed 
$R$~= 4.2~$\pm$~0.2 and Vrba \etal\ (1993\markcite{VCT93}) determined 
$R$~$\simeq$~4 in the nearby $\rho$ Ophiuchi dark cloud.  Peterson 
\etal\ also estimated an $R$~$\sim$~4 for this cluster.  Many other 
investigations, with similarly high $R$ results, are reviewed by Dixon 
\& Longmore (1993).

To examine this issue, we used our \teff\ values to also estimate 
intrinsic M4 $V-K$ colors, by again interpolating the model atmosphere 
predictions of Cohen \etal\ (1978\markcite{CFP78}).  We compared these 
predicted values against the observed $V-K$ colors of Frogel \etal\ 
(1983, their table 17A) after removing the reddening corrections they 
applied:  Frogel \etal\ (1983) had adopted the ``classical'' values of 
$E(V-K)$~= 2.78$\times E(B-V)$.  By subtracting the modeled 
colors from the observed photometry for the thirteen stars in common 
between our study and theirs, and adding one more star to the list 
using the $B-V$ photometry of Cudworth \& Rees (1990\markcite{CR90}) 
along with the $V-K$ photometry of Fusi Pecci \& Ferraro 
(1997\markcite{FF97}), we determined color excesses $E(V-K)$ and 
$E(B-V)$ for fourteen stars in M4, and derived a value of $R$~= 
3.4~$\pm$~0.4, in reasonable agreement with those of previous M4 
studies.

\subsection{Abundance Results}

Most abundances were derived as described above, from single-line 
analyses matching predicted and measured EWs.  In addition to employing 
blended-line EW computations for \nion{V}{i} lines, we also used these 
for determining Ba abundances.  The \nion{Ba}{ii} lines have both 
hyperfine and isotopic subcomponents, and for these we used the line 
lists of McWilliam \etal\ (1995\markcite{MPSS95}).  Solar abundance 
ratios among the $^{\rm 134-138}$Ba isotopes were assumed in these 
calculations.  In view of our subsequent determinations of large [Ba/Eu] 
ratios, either a solar system isotopic distribution or one more heavily 
weighted toward those isotopes generated by $s$-process neutron-capture 
synthesis (which would {\it increase} the Ba overabundances) seems to be
appropriate.  Finally, full synthetic spectrum computations were used to 
derive abundances from the $\lambda\lambda$6300, 6363~{\rm \AA}\ 
[\nion{O}{i}], the \wave{5711} \nion{Mg}{i}, and the $\lambda\lambda$5682 
and 5688~{\rm \AA}\ \nion{Na}{i} lines.  Comments on most of the abundance 
computations appear in earlier papers in this series and will not be 
repeated here.  See \S 3.6 for additional discussion of [\nion{O}{i}] line 
analyses.

In Table~\ref{Ivans.tab5} we list the final abundances for each star 
and the cluster mean abundances.  Figure~\ref{Ivans.fig7} presents a 
``boxplot'', useful for exploratory data analysis, to summarize the mean 
and scatter of each element.  This boxplot illustrates the median, data 
spread, skew and distribution of the range of values we derived for each 
of the elements from our program stars.  The boxplot also clearly 
illustrates possible outliers.  For example, the abundance ratio range we 
obtain for proton-capture elements such as sodium and oxygen is quite 
large.  However, the star-to-star abundance variations are very small for 
all the heavier elements, and for only Ti and V has this result been 
somewhat pre-ordained.  

Table~\ref{Ivans.tab6} shows the results of numerical experiments to 
determine the sensitivity of derived elemental abundances to model 
parameter changes (i.e. potential errors; \S3.3). We explored the effects 
in L2206, a star of intermediate temperature and \logg\ for which we found 
\teff/\logg/\vmicro\ = 4325/1.35/1.55. The effects of both random and 
possible systematic \teff\ errors are shown.  For most species, the model 
uncertainties induce relative abundance errors of less than $\pm$0.05~dex.  
The abundances derived from the \nion{Fe}{i} lines have somewhat larger 
sensitivities to microturbulent velocity uncertainties, the \nion{Fe}{ii} 
and \nion{Mg}{i} lines have larger sensitivities to the derived gravities, 
and the \nion{Ti}{i} and \nion{V}{i} lines have the largest sensitivities 
to the derived \teff.

Figure~\ref{Ivans.fig8}  illustrates the run of Fe-peak and 
neutron-capture element abundances with \teff.  The scatter about the 
mean (of elements not expected to be sensitive to proton-capture 
nucleosynthesis) compares well to those obtained in other high resolution 
cluster work and, as expected, no trend with \teff\ is observed for these 
non-volatiles.  The abundances of these elements have no apparent trends 
with RGB or AGB position, since there are no perceptible slopes with 
\teff\ depicted in Figure~\ref{Ivans.fig8}.  Sneden \etal\ (1997) have 
found large ($\sim$~0.4) scatter in [Ba/Fe] and [Eu/Fe], with 
$<$[Ba/Eu]$> \sim$~--0.4 among 18 M15 giants, but that obviously is not 
seen in our M4 sample, in which $<$[Ba/Eu]$>$~= +0.60~$\pm$~0.02 
($\sigma$~=0.10); the scatter here is probably dominated by 
observational/analytical uncertainties.

Usually, calcium abundances are sturdy enough to also be included in a 
plot such as Figure~\ref{Ivans.fig8}. However, the scatter in the 
derived calcium abundance ($\sigma$ = 0.11) is a little larger than that 
expected due to observational error alone and demands further 
investigation.  Previous work using similar analysis techniques on data
of comparable resolution and sample size include those of M13 by Kraft 
\etal\ (1997\markcite{KSSSLP97}) and M15 by Sneden \etal\ 
(1997\markcite{SKSSLP97}) for which 18-star samples of calcium abundances 
show $\sigma$ = 0.03 and 0.05, respectively. Simply eliminating calcium 
lines that had not been included in previous Lick-Texas cluster analyses 
did not decrease the variation in the abundance results.   We also 
investigated the possibility that uncertainties in EW measurements, 
combined with our choices in microturbulent velocities, could result in 
the derived calcium abundance variations.  However, seven of the calcium 
lines we use are in common with those of Drake \etal\ 
(1994\markcite{DSS94}), who studied four stars in common with our sample. 
For the 18 calcium line measurements in common between the two studies
(where the average EW is~$\sim$~85~m{\rm \AA}), 
we find a mean difference in the sense {\it this study minus Drake et 
al.} of 3.3 $\pm$ 1.4~m{\rm \AA}.  And, although our mean abundance for 
these stars is slightly offset from their study (Drake \etal\ adopted 
systematically higher surface gravity values but found similar 
microturbulent velocities), the relative range of difference in the 
four-star sample is approximately the same, with ours slightly more 
restricted (we find a total calcium abundance range of 0.08 for these 
four stars and Drake \etal\ find 0.13).  We also compared our abundance 
results for the three giant stars we studied in common with Brown \& 
Wallerstein (1992\markcite{BW92}): we find a mean difference in the sense 
{\it this study minus Brown \& Wallerstein} of 0.02~dex, with two of the 
stars showing much lower than average Ca abundances in both studies.

M4 stars are known to have a range of CN-strengths.  We find that the 
scatter in our calcium abundances seemed to correlate with the
CN-strength of our stars: the calcium difference, in the sense of 
{\it [Ca/Fe]$_{CN-Strong}$ -- [Ca/Fe]$_{CN-Weak}$} shows a large scatter, 
0.08 $\pm$ 0.11~dex.  Drake \etal\ (1994\markcite{DSS94}) also found 
that their ``{\it calcium results suggest that this element might be 
slightly more abundant in the CN-strong stars than in the CN-weak stars}''.  
They investigated the possibility that stronger calcium lines may be the 
result of cooler outer envelopes that have large cyanogen opacities.  
They found that atmospheric temperature effects due to differential 
molecular blanketing are not responsible for abundance variations of Na, 
Al, O, {\it or} Ca.  While upper atmosphere cooling {\it can} produce 
correlations between Na and Al abundances with CN-strengths, a 
corresponding correlation would be expected in the resonance lines of 
neutral potassium which is even more sensitive to upper atmosphere 
cooling (\nion{K}{i} has a lower ionization potential than either 
\nion{Na}{i} or \nion{Al}{i}). However, no such correlation was found 
for the two pairs of CN-strong and CN-weak M4 stars investigated by 
Campbell \& Smith (1987\markcite{CS87}).  While none of their stars 
overlap our high resolution sample, we are able to verify, using the 
$\lambda$~7699~{\rm \AA} \nion{K}{i} line which appears in a 
telluric-free region redward of the A-band on our program star spectra, 
that there is no difference in \nion{K}{i} strength between the CN-strong 
and CN-weak pairs in our sample.  

Finally, we tested whether the scatter in the Ca abundance might be due 
to an inability of our EW analysis to properly account for CN-blending in 
the spectra of the CN-strong stars: we looked for the problem to appear 
in EW analyses of other elements {\it known} to be affected by CN, such 
as oxygen.  However, a comparison of the scatter in the oxygen abundance 
results between those obtained by the EW analysis procedure ($\sigma$ = 
0.15) with those obtained by spectrum synthesis ($\sigma$ = 0.14) shows 
no significant difference in the dispersion of values obtained by both 
methods and reveals only the expected offset due to the proper accounting 
of CN-blending in the synthesized spectrum.  Despite the preceding tests, 
we are not satisfied that the scatter is of nucleosynthetic origin:  most
of AGB stars are also CN-weak stars (this is discussed further in \S4.2.3) 
and we note that a 50K upward revision in temperature (see 
Table~\ref{Ivans.tab6}) would increase the Ca abundance in these stars to 
match that of the CN-strong RGB stars.

In Table~\ref{Ivans.tab7} we restate our mean M4 abundances, 
combining the results for the two ions of Fe and Ti as straight averages.  
We also tabulate the abundances of Gratton \etal\ (1986), Brown \& 
Wallerstein (1992), and Drake \etal\ (1992).  For contrast, we also list 
mean abundances in M5, in typical field metal-poor stars, and also the 
very metal-poor cluster M15.  For this simple overview we chose not to 
break down the comparisons on a star-by-star basis; only cluster means 
are considered.  Even with this limitation, it is clear that our M4 
abundances generally agree well with those of past M4 investigators.  
The M4 abundances of non-volatile elements Ca, Sc, Ti, V, and Ni also are 
in good accord with those in M5 and the halo field (again with the 
warning that Ti and V abundances have been partially forced to this 
conclusion).  The $r$-process neutron-capture element Eu also has 
essentially the same mean abundance in M4, M5, M15, and the field stars.  
Our M4 average Na and Si abundances are substantially larger than those 
of M5 and the halo field stars but agree with those of the very 
metal-poor cluster M15.  But all M4 analyses agree on this point.  The 
elevated Na and Si abundances will be explored further in \S5

The only M4 abundance disagreement substantially in excess of 
observational uncertainties (typically $\sim$~0.1 in [el/Fe] ratios) is 
with the Gratton \etal\ (1986) Ba abundance.  Both we and Brown \& 
Wallerstein derive [Ba/Fe]~$\sim$~+0.6, while Gratton \etal\ suggest that 
this ratio is $\sim$~0, about the solar system ratio.  This single 
discrepancy cannot be resolved here, as we have no high resolution 
spectra for the three M4 stars (L1605, L2608, and L2626) of the Gratton 
\etal\ sample.  But, if we and Brown \& Wallerstein (1992) are correct 
about the mean Ba abundance in M4, this element is extremely overabundant 
with respect to the comparison samples.

Some comment must be made on the reliability of the Ba and Si abundances
in our study.  Not only do our results for Ba agree with those of Brown 
\& Wallerstein,  they are also in accord with those derived for quite 
different M4 stars by Lambert \etal\ (1992\markcite{LMS92}).  Lambert 
\etal\ obtained [Ba/Fe] = +0.50 and +0.63 in two blue HB stars, using 
lines of intermediate strength.  However, we caution the reader that our 
Ba abundances are derived from very strong lines and thus are extremely 
sensitive to adopted microturbulent velocities.  We can lower the [Ba/Fe] 
by $\sim$~0.3~dex simply by increasing our microturbulence velocities by 
a very large amount: $\sim$~1.0~\kmsec. But, doing so produces 
unforgivably large trends of \nion{Fe}{i} line abundances with EWs.  The 
\vmicro\ we chose for each star was in accord with the abundance 
indications we used to determine the final \teff\ and \logg\ parameters 
we adopted for the models.  Our estimate of the random error of the 
microturbulence velocities, based on the dependence of \nion{Fe}{i} line 
abundances with EW, is 0.15~\kmsec, too small an error to explain away 
the very large Ba abundances we derived.  However, without \nion{Ba}{i} 
lines in our spectrum or weaker \nion{Ba}{ii} lines for which the atomic 
parameters are well known, there is no resolution to the dilemma at this 
time.  We hope to resolve the issue in a future investigation by 
comparing this element, and other $s$-process element abundances, against 
those derived for M5 stars of similar \teff, \logg, and metallicity 
observed with the Keck HIRES.  

We have confirmed the apparent excess of $s$-process species in M4 using 
the relatively weaker lines of lanthanum.  We derived \nion{La}{ii} 
abundances using the CN-free line at $\lambda$6774.33{\rm \AA} and, 
especially in the cases involving the hotter, higher \logg\ stars (where 
the line was very weak), checking to ensure that the result was in 
reasonable accord with results derived from $\lambda\lambda$ 5808.31 and 
6390.49{\rm \AA} (CN-blended lines).  Although literature values of La 
abundances are sparse for metal-poor stars, Table~\ref{Ivans.tab7} shows 
both the excellent agreement we obtain with previous work in M4 (Brown \& 
Wallerstein 1992\markcite{BW92}) as well as the overabundance of La as 
compared with field halo giants of comparable metallicity.

While our silicon abundance results are in good agreement with previous 
studies of M4, they are significantly higher than the abundances found 
either in M5 or in the field.  Unlike the \nion{Ba}{ii} lines used in our 
analysis, the \nion{Si}{i} lines are all of intermediate or weak 
strength, with very little sensitivity to random errors in the 
fundamental parameters of the stellar models.  Comparing our EW measures 
of the \nion{Si}{i} lines against those made available by J. Brown (1998), 
the EWs of these lines show the same average offset and scatter as the 
rest of the sample.  However, high Si abundances are seen in the very 
metal-poor cluster M15 and are thus not unprecedented among globular 
clusters.  Further exploration of this issue is postponed to \S5.

\subsection{Carbon and Nitrogen Abundances}

Our M4 spectra did not contain spectral features suitable for 
independent determination of carbon and nitrogen abundances.  But, by 
making use of previous investigations of these elements we were able to 
compute total abundances of the CNO group.  Since the [\nion{O}{i}] 
$\lambda\lambda$6300, 6363${\rm \AA}$ lines are blended with CN, we also 
explored the effects of the adopted C and N abundances on the derivation 
of oxygen abundances.  Finally, C and N abundances were also employed in 
deriving \carbiso\ ratios, using the CN red system features near 
8000~{\rm \AA}.

Suntzeff \& Smith (1991\markcite{SS91}, hereafter SS91) used low 
resolution (R~$\sim$~2000) CTIO IRS spectra of first-overtone 
vibrational bands of CO near the 2.3$\mu$m to determine C abundances for 
32 M4 giants.  With only CO band observations at hand, their derived C 
abundances were not formally independent of assumed O abundances.  They 
assumed a uniform oxygen overabundance of [O/Fe]~=~+0.3 in synthetic 
spectrum computations of the CO bands.  This is slightly larger than the 
mean abundance found in our study: from Table~\ref{Ivans.tab5}, 
$<$[O/Fe]$>$~= +0.25~$\pm$~0.03 ($\sigma$~=~0.14).  However, since C is 
always at least a factor of 10 less abundant than O in M4 giants, the CO 
strength is almost entirely controlled by the available C atoms.  SS91 
did explore the effect of reducing their assumed O abundance by 0.3~dex 
and found that the derived C abundance changed by less than 0.1~dex.  
Thus no adjustment in their C abundance scale was deemed necessary for 
this potential source of uncertainty.

We paid more attention to model atmosphere parameter differences with 
SS91.  We have 12 stars observed at high resolution that are in common 
with their work.  They assumed lower temperatures, higher gravities, as 
well as higher microturbulent velocities than we did.  Employing our 
fundamental parameters would yield systematically higher abundances;  
the biggest C abundance effect comes from a combined temperature/gravity 
shift.  We therefore adjusted their C abundances to our temperature 
scale by doing a least squares fit in \teff\ to their data set (excluding 
one large outlier) using standard linear regression techniques, and used 
this function to calculate the difference in assumed C as a function of 
temperature.  In Table~\ref{Ivans.tab8}, we tabulate the SS91 C 
abundances and the adjusted values, and in Figure~\ref{Ivans.fig9} we plot 
their C abundances as a function of \teff\ for 29 stars, excluding the 
anomalous L1412.  The C abundances tend to decrease with decreasing 
\teff; this probably is the result of increased mixing of CN-cycle 
material to the stellar surfaces as M4 stars ascend the RGB.  For the 12 
stars in common with us (solid circle symbols) we also show arrows to the 
adjusted ${\rm log}~\epsilon$(C) values at the higher \teff's derived in 
this study.

Using these C abundances, we calculated synthetic spectra for portions 
of the CN red system $A^{2}$II--$X^{2}\Sigma^{+}$ (2--0) band near 
\wave{8000}, and derived N abundances.  For the synthetic spectra we 
adopted the atomic/molecular line lists of Gilroy (1989).  In 
Table~\ref{Ivans.tab8} the N abundances determined in this manner are 
given, and the O abundances for all of our ``high-resolution'' stars 
from Table~\ref{Ivans.tab5} also are retabulated here in 
${\rm log}~\epsilon$(el) form.

The derived mean M4 giant star C abundance is 
$<{\rm log}~\epsilon$(C)$>$~= 6.81$~\pm$~0.07 ($\sigma$~=~0.25), or
$<$[C/Fe]$>$~= --0.57~$\pm$~0.07,\footnote{Here we have assumed the solar 
CNO abundances of Anders \& Grevesse (1989): ${\rm log}~\epsilon$(C)~=
8.56, ${\rm log}~\epsilon$(N)~= 8.05, and ${\rm log}~\epsilon$(O)~= 8.93.} and 
the derived mean N abundance is 
$<{\rm log}~\epsilon$(N)$>$~=~7.62~$\pm$~0.13 ($\sigma$~=~0.47), or 
$<$[N/Fe]$>$~= +0.75$\pm$~0.13.  Our results compare reasonably well with 
two other studies of C and N in M4. Brown \etal\ (1990\markcite{BWO90}), 
employing spectrum synthesis of CH features of the $A-X$ system in the 
blue wavelength region, obtained $<{\rm log}~\epsilon$(C)$>$~= 
6.97~$\pm$~0.06 for their three non-peculiar stars.  Adopting this C 
abundance, they then obtained their N abundance of 
$<{\rm log}~\epsilon$(N)$>$~= 7.83~$\pm$~0.15 by doing an equivalent 
width analysis of (2--0) band lines of the CN red system.  Smith \& 
Suntzeff (1989\markcite{SS89}), using low resolution (R~$\sim$~2000) KPNO 
FTS spectra taken of first-overtone bands of CO in the $K$-band near 
2.3$\mu$m, derived the C abundance for seven M4 stars, six of which are 
in common with this study.  Excluding the peculiar star L1412, the 
average C abundance they obtained is 
$<{\rm log}~\epsilon$(C)$>$~= 6.75~$\pm$~0.23.  

The C+N+O abundance sums are also given in this table.  They are constant 
to within the analysis uncertainties, with a mean value of 
$<{\rm log}~\epsilon$(C+N+O)$>$~= 8.24~$\pm$~0.03 ($\sigma$~=0.09).  This 
is just the total C+N+O one would expect if the material out of which the 
M4 giants formed was enhanced in O over the scaled solar abundance by 
[O/Fe]~$\simeq$~+0.4, as is expected from previous work on metal-poor 
globular clusters.   

The derived or extrapolated C and N values were then adopted in the 
$\lambda\lambda$6300, 6363${\rm \AA}$ regions where the oxygen abundances 
were determined by spectrum synthesis.  Some trials were made, adjusting 
the C and N abundances up and down but keeping the total of [C/Fe] plus 
[N/Fe] abundance (hereafter C+N) constant.  The oxygen abundance, at 
least to the level of accuracy that can be determined with these data, 
did not depend on the relative ratios of [C/Fe] to [N/Fe], just the total 
C+N.  This is because the CO association is relatively small in the CN 
line-forming layers of the M4 RGB stars, and thus C is mostly in the 
neutral atomic state.  A decrease in either C or N abundance therefore 
must be compensated by an increase in the other atom in order to keep CN 
line strengths approximately constant.  Other numerical experiments we 
performed showed that changing the C or N abundance alone, and holding 
the other values fixed, made less than one percent difference in the 
depths of the modeled [\nion{O}{i}] profiles.

\subsection{The \carbiso\ Ratio}

We also determined carbon isotope ratios \carbiso\ for our program stars. 
The dissociation potentials of CN and CO molecules are not sensitive to 
isotopic substitution.  Thus, we assumed that the number-density ratio of 
$^{12}$CN to $^{13}$CN molecules was equal to that of the isotopic 
abundance ratio \carbiso.  The ratios were determined using spectrum 
synthesis of portions of the (2--0) band of the CN red system near 
\wave{8005}.  Once the N abundance had been determined from $^{12}$CN 
features for a constant assumed isotopic ratio, the nitrogen abundance 
was fixed and the ratio was allowed to vary.  A best fit to the data was 
determined by eye.  The derived \carbiso\ value for each program star is 
given in Table~\ref{Ivans.tab8}, followed in parentheses by the minimum 
and maximum possible values that the fitting procedure would allow.  The 
average ratio obtained for our program stars is 
$<$\carbiso$>$~=~4.5~$\pm$~0.1 ($\sigma$~=~0.6).

The \carbiso\ ratio, based on a comparison of line intensities 
originating in similar or identical rotational and vibrational levels, 
is insensitive to errors in adopted effective temperatures, surface 
gravities, line oscillator strengths, and molecular dissociation 
energies (see the discussion in Lambert \& Sneden 1977\markcite{LS77}), 
making it a robust indicator of the presence of CNO-cycle processed 
material. We investigated the effect on the abundance ratio resulting 
from possible errors in our N abundance determinations only to verify 
that the iterative approach we used to obtain the isotope ratio and 
assumed N abundance is a self-correcting one, assuming some fixed value 
for the C abundance.

Our results fall within the range of values found by Brown \& 
Wallerstein (1989\markcite{BW89}) and SS91 for the stars in common.  
This excellent agreement with previous studies confirms from CN red 
system syntheses the extremely low carbon isotope ratios found from the
CO bands.  For the evolutionary domain of our program stars, there is no 
obvious trend of \carbiso\ ratios as a function of giant branch 
luminosity.

\subsection{The CN-Strong, CN-Weak Dichotomy: A New CN-Strength Index}

M4 has a well-developed CN-strong, CN-weak dichotomy which we propose 
to discuss in relation to the carbon, nitrogen, and oxygen abundances 
derived from the present observational material.  We observed 14 stars 
at high resolution in common with either Norris (1981\markcite{Nor81}) 
or SS91\markcite{SS91}, who determined CN strengths for these stars 
using S(3839) indices (measures of the ratios of the flux intensities 
of the cyanogen band near 3839~{\rm \AA}). However, this left ten stars 
in our high resolution sample for which a CN classification was unknown.  
While the signal-to-noise of our spectra in the red is quite good, the 
flux in the blue is too poor to independently determine either the 
carbon abundances from the CH G-band, or the S(3839) CN-strength index.  
In order to investigate the dependence of other light element abundances 
on the CN strengths of all our stars, we developed a simple CN-strength 
measure from the CN (2--0) bandhead at 7872~{\rm \AA}.  In 
Figure~\ref{Ivans.fig10} we show the appearance of this bandhead in M4 
program stars with varying amounts of CN absorption.  We determined a 
``bandhead'' EW via a straight numerical integration of the total 
absorption in the wavelength region $\lambda$7872.4--7876.4~{\rm \AA}.  
This EW, which we call EW7874, covered several CN lines, and this 
provides a measure of the CN-strength for the stars in our high 
resolution sample.  

Figure~\ref{Ivans.fig11} shows our EW7874 measure as well as the 
S(3839) index as determined by either Norris or SS91, as a function of 
$V$ magnitude.  The baseline for the S(3839) plot is from Norris.  As 
illustrated in the figure, there is a variation in the CN-strength with 
magnitude, regardless of which measure is used.  Stars on the giant 
branch have correlated color and magnitudes: some of the CN-strength we 
measure in EW7874 arises from temperature effects, with the cooler stars 
showing stronger molecular features in general.  However, we were unable 
to independently determine the \teff\ baseline for EW7874 using our 
relatively smaller sample of stars.  Instead, the \teff\ baseline for 
EW7874 was obtained by taking advantage of the previous work done by 
Norris and SS91, adopting the linear slope in \teff\ which provided the 
best resulting correspondence with the previous designations for those 
stars we observed in common.  Then, the relative CN-strength values for 
all stars were determined as offsets from this line:
$$
\delta{\rm EW7874} \equiv {\rm EW7874} - (165.784 - 0.025\times\teff).  
$$
The values $\delta$EW8984, our adopted parameter of the CN strength for
all of our high resolution stars, as well as the published values of 
S(3839) for the stars in common, are given in Table~\ref{Ivans.tab8}.
We find only two mildly discrepant points (L1408 and L2206), for which 
we determine intermediate CN-strengths but Norris and/or SS91 designate 
as strong, according to their blue CN index.  Both stars are among those 
in our R$\sim$~60,000 sample; continuum placement is not a problem.  
Telluric lines in the EW7874 region are shown in 
Figure~\ref{Ivans.fig10} and one would require a much too large 
over-division of the telluric lines in order to explain the weakness of 
the red index.  However, the indices for the remainder of the stars in 
common are in good agreement.

By placing the red EW7874 estimate, in effect, on the S(3839) system, 
we are now able to give CN-strength designations to all of our stars 
observed at high resolution: we will denote stars as either CN-weak 
($\delta$EW7874 $\le$ 64) or CN-strong ($\delta$EW7874 $\ge$ 86) in the 
remaining plots that follow in our discussion.  But here we alert the 
reader to two caveats about this scheme. First, adoption of this two-bin 
CN-designation deliberately avoids finer grouping distinctions of stars 
with intermediate CN strengths.  For the purposes of this discussion, 
the coarser CN-strong/CN-weak notation will suffice. Second, we do not 
have an equal number of CN-weak and CN-strong stars.  To assume so would 
have forced us to choose a lower EW7874 value for the division. Of the 
stars with the highest $\delta$EW7874 values we have designated as 
CN-weak, the spectra are all similar in appearance and to distinguish, 
by number alone, CN-strong versus CN-weak among these stars would be 
misleading.  Instead, we have taken advantage of the previously 
determined CN-strengths assigned from the S(3839) system, as well as 
the distinctions we can discern both by eye and by measure in the 
EW7874 region, rather than make the designations blindly, by raw number 
alone.

\subsection{Medium Resolution Abundances}

We determined only oxygen abundances from the medium resolution Argus 
spectra (Table~\ref{Ivans.tab9}).  We first analyzed the stars in 
common with the high resolution observations, using the previously 
determined \teff, \vmicro, \logg, and overall metallicity [Fe/H] values.
In \S2.2 we explained the procedures used to eliminate the major 
telluric O$_{\rm 2}$ contamination that plagues the \wave{6300} region 
of the Argus spectra.  During the O abundance analysis this issue was 
revisited, and the results of the [\nion{O}{i}] syntheses were tested 
for the effect of different choices in telluric line division.   Since 
the Argus multi-object spectrograph observed all of the program stars at 
the same time through the same air mass, after these numeric experiments 
we applied the telluric line removal procedure in a consistent manner to 
all Argus stars.  Despite the telluric line removal complication caused 
by the known resolution change across the fibers (\S2.2), the agreement 
of O abundances between the medium and high resolution data sets is 
excellent.  Taking the difference in the sense {\it high resolution 
minus medium resolution}, $<\delta$[O/Fe]$>$~= --0.01~$\pm$~0.02 
($\sigma$~= 0.07, for 13 stars in common).

For the remaining eleven stars, we adopted the mean metallicity of M4
([Fe/H]~= --1.18) and a microturbulent velocity \vmicro~=~1.75~\kmsec. 
This \vmicro\ is the mean of the warmer stars observed at high 
resolution with \teff~$>$~4325~K (a temperature no cooler than the 
lowest predicted \teff\ of the medium resolution sample).  In deriving 
the stellar temperatures, we again utilized the ``Gray'' method, but 
calibrated the ratios using the results we obtained using the high 
resolution spectra.  The wavelength coverage of the medium resolution 
data includes five of the lines we had previously measured and used to 
derive the initial temperature rankings of the high resolution data.  At 
this resolution however, all of the lines were blended, and two original 
Gray ratios proved to be useless for these data and were eliminated.  
The remaining three lines, not as severely blended, formed two pairs; we 
measured the line depths for these lines.  The logarithmic line depth 
ratios were fit as a second-order polynomial function to the 
corresponding temperatures and the inverted form of this function was 
used to predict the temperatures of the remaining stars.  For only one 
star (L1617) was an extrapolation required based on its depth ratio.  We 
calculated an evolutionary gravity for each star, adopting the distance 
(2.1~kpc) and extinction ($A_V$~=~1.48) that had produced the best match 
between the spectroscopic and evolutionary gravities found earlier from 
the high resolution data.  We adopted cluster mean abundances in the 
synthetic spectrum computations, and resulting [O/Fe] values for these 
11 medium-resolution-only stars are given in Table~\ref{Ivans.tab9}.

\section{M4 Nucleosynthesis}

In this section we examine the probable causes of the abundance trends 
found in our study.  We first summarize the main observational results 
emerging from the previous sections.  From this large data set we have 
found the following:

\begin{enumerate}

\item The Fe metallicity of M4 is $<$[Fe/H]$>$~= --1.18 
($\sigma$~=~0.02) for the 23 M4 stars observed at high resolution, 
excluding the pathological star L1412 and the possible binary L2406 
(see \S4.1). The dependence of the metallicity on analysis assumptions 
has been discussed in \S3; this probably adds an additional scale 
uncertainty of less than $\pm$0.10 to the scatter about the mean quoted 
here.

\item The deduced reddenings, total extinctions, and cluster distance
are reasonably consistent with previous investigations.  Our analysis 
confirms the earlier claims of star-to-star extinction variations, with 
a generally increasing trend seen across the face of the cluster, from 
east to west.  The total-to-selective extinction ratio is larger than 
the typical ISM value, again in agreement with the work of prior 
investigators.

\item Abundance ratios with respect to Fe typical of halo field and
cluster stars are obtained for Sc, Ti, V, Ni, and Eu.  The star-to-star 
abundance variations in all these elements are $<$$\pm$0.1 (from the 
$\sigma$ values in the [X/Fe] ratios).

\item Very high relative abundances for Si and Ba are obtained in our
study ([Si~or~Ba/Fe]~$\sim$~+0.6), as well as high La 
([La/Fe]~$\sim$~+0.5) which echo the results of the earlier investigation 
by Brown \& Wallerstein (1992).  High Si abundances, although not common 
among globular cluster giants, have also been found in the very metal-poor 
cluster M15.

\item Star-to-star variations in the proton-capture elements O, Na, and 
Al are detected.  Al is overabundant in all M4 giants: 
$<$[Al/Fe]$>$~$\gtrsim$~+0.4.

\item Substantial enhancement of N abundances is deduced
($<$[N/Fe]$>$~$\sim$~+0.7) that accompanies often large C depletions 
($<$[C/Fe]$>$~$\sim$~--0.6).  The carbon isotope ratio is also extremely 
low in all stars of our sample.  Although several M4 giants exhibit 
oxygen deficiencies, most M4 giants show little evidence for severe O 
depletions ($<$[O/Fe]$>$~$\sim$~+0.25), such as are seen in M13 (Kraft 
\etal\ 1997\markcite{KSSSLP97}) and M15 (Sneden \etal\ 
1997\markcite{SKSSLP97}).  Excluding the two coolest program stars (for 
which we place large uncertainties on our abundance results), only three 
stars have [O/Fe]~$\leq$~+0.10.  With the caution that the derived C, N, 
and O abundances have substantial analysis uncertainties, we conclude 
that the C+N+O abundance sum is constant to within the observational 
errors, and agrees with the C+N+O total that might be expected for M4 
stars at birth.
\end{enumerate}

\subsection{An H-R Diagram for M4 Giants}

Interpretation of our abundance data requires knowledge of the 
evolutionary states of the M4 program stars.  The variably-reddened 
($V,B-V$) c-m diagram of Figure~\ref{Ivans.fig1} is clearly inadequate 
for this task, so in Figure~\ref{Ivans.fig12} we display an 
($M_{bol}$,\teff) H-R diagram whose quantities are based on our 
spectroscopic results.  The abscissa \teff\ values are those of
Table~\ref{Ivans.tab2}.  The ordinate $M_{bol}$ values have been 
computed from the $V$ magnitudes of Table~\ref{Ivans.tab1}, derived 
$E(B-V)$'s from Table~\ref{Ivans.tab4}, bolometric corrections BC from 
Worthey (1994), and the true M4 distance modulus $(m-M)$~=~11.61 (from 
our preferred distance $d$~=~2.1~kpc).  For these choices of $d$ and 
$A_V$, the brightest M4 stars have $M_{bol}$ $\sim$~--3.5, a value 
which agrees well with the theoretical $M_{bol}$ ($\sim$~--3.55) 
expected at the He core flash ({\it e.g.} Rood 1972\markcite{Roo72}, 
Sweigert \& Gross 1978\markcite{SW78}).

In Figure~\ref{Ivans.fig12} the RGB tip is clearly defined, as is 
the AGB split from the lower RGB.  Note that a comparison of 
Figures~\ref{Ivans.fig1} and \ref{Ivans.fig12} emphasizes that the 
number of M4 AGB stars {\it observed} in our survey overrepresents 
their true numbers relative to lower RGB stars.  This is entirely an 
artifact of our magnitude-limited high resolution sample.  The defect 
is partially repaired in our medium-resolution sample, which has about 
eight lower RGB stars.  For nearly all program stars, the derived H-R 
diagram illustrates far less scatter than the observed c-m diagram.

One anomalous star stands out in Figure~\ref{Ivans.fig12}: L2406,
alias V13.  It is variable in spectral type (G3--K0, Joy 
1949\markcite{Joy 1949}), and in $V$ magnitude (12.4--13.1, Sawyer 
Hogg 1973\markcite{SH73}; see also Lloyd Evans 1977\markcite{LE77}).  
Our analysis yielded a (\logg,\teff) combination which, when convolved 
with the observed $V$ magnitude and our predicted $E(B-V)$, produces a 
star with the anomalously bright $M_{bol}$ depicted in 
Figure~\ref{Ivans.fig12}.  The appearance of the L2406 spectrum is 
in fact very similar to two other fainter stars we found in the same 
\teff~$\simeq$~4100~K range, but the microturbulent velocity proved to 
be unusually high.  Closer inspection of the spectrum revealed that its 
absorption lines are much broader than those of other M4 giants, and 
are slightly blueward asymmetric.  It is thus possible that L2406 is a 
close physical (or line-of-sight) binary consisting of two giant stars 
of similar magnitude and temperature.  A binary interpretation would 
not itself explain the photometric and spectral type variability of 
this star, but a comprehensive investigation of the nature of L2406 is 
beyond the scope of this paper.  The star will be excluded from further 
discussion and interpretation of our M4 abundance results.

\subsection{Tests of Proton-Capture Nucleosynthesis}

Armed with a rectified M4 upper giant branch H-R diagram and 
CN-strength indices for essentially all of the stars from our high 
resolution sample, we investigate the abundances of those elements 
that are sensitive to proton-capture nucleosynthesis (via the CN-, 
ON-, NeNa- and MgAl-cycles): carbon, nitrogen, oxygen, sodium, 
magnesium and aluminum.  We consider the relationship of these 
abundances first to each other and then to evolutionary state.

\subsubsection{C, N, and O}

Figure~\ref{Ivans.fig13} illustrates the relationship between C and 
N versus O for the 12 stars of Table~\ref{Ivans.tab8} for which all 
three elements have abundance measurements (recall from \S3.3 that 
the open square symbol, $\Box$, here and in subsequent figures, 
refers to a high luminosity, low \teff\ star for which model 
parameters and abundance determinations are relatively uncertain).  
The distribution of points is compatible with a proton-capture 
scenario:  low oxygen abundances are accompanied by low carbon and 
elevated nitrogen.  The more highly CNO-processed stars are usually 
also the CN-strong stars.  In addition, the sum of C+N+O is 
essentially constant, as expected, if all stars draw on the same 
primordial material. Note also the existence of one star (L4201, 
observed at slightly lower resolution, R $\sim$~30000, than the bulk 
of our high resolution sample), in which O appears to have remained 
undepleted, whereas C$\rightarrow$N conversion has taken place, which 
suggests that the material of this star has been subjected to 
less-advanced nucleosynthetic processing than has been the case for 
the other stars of our sample.

\subsubsection{Na and Al {\it vs.} O; Mg and Al {\it vs.} Na}

In Figure~\ref{Ivans.fig14}, we plot sodium and aluminum abundances 
{\it vs.} oxygen.  The anticorrelated behavior of Na relative to O, 
common among giants in many other globular clusters ({\it e.g.} 
Suntzeff 1993\markcite{Su93}, Briley \etal\ 1994\markcite{BBHS94}, 
Kraft 1994\markcite{Kr94}, Wallerstein \etal\ 1997\markcite{Wetal97}), 
is clearly seen. Less striking is the apparent anticorrelated behavior 
of Al and O. These anticorrelations exhibit some of the characteristics 
of the proton-capture deep-mixing scenario (Langer \etal\ 
1993\markcite{LHS93}, Langer \& Hoffman 1995\markcite{LH95}, Langer 
\etal\ 1997\markcite{LHZ97}, Cavallo \etal\ 1998\markcite{CSB98}) in 
which Na and Al are enhanced in the stellar envelope at the expense of 
Ne and Mg, respectively, as the envelope is circulated through the 
hydrogen-burning shell.  As is seen in Figure~\ref{Ivans.fig13}, the 
CN-strong stars are those that are more highly processed via 
proton-capture syntheses.  The CN-strong group has a mean Na abundance 
that is a factor of two larger than the CN-weak group: $\Delta$[Na/Fe] 
= 0.32 $\pm$~0.06.  Correlated Na and CN variations are also observed 
in a few pairs of stars in NGC~6752 (Briley \etal\ 
1999\markcite{BSSBHLS99}), a cluster of comparable metallicity to that 
of M4.  Our CN-strong group also has higher Al abundances but the 
CN-strong/CN-weak difference is much less pronounced: $\Delta$[Al/Fe] =
0.16 $\pm$~0.05.  This situation is the reverse of that found in M13, 
in which the range in Al abundances far exceeds that of Na (Kraft \etal\ 
1997\markcite{KSSSLP97}).

Two stars stand out which do not completely follow the trends exhibited
in Figure~\ref{Ivans.fig14}: L4201 and L2208. We noted already that 
L4201 has an anomalous position in Figure~\ref{Ivans.fig13}, which 
indicates that, whereas C has been converted to N, O remains unchanged. 
The fact that Na is enhanced in this star whereas Al remains relatively 
low is consistent with theoretical predictions (Langer \etal\ 
1997\markcite{LHZ97}, Cavallo \etal\ 1998\markcite{CSB98}) which argue 
that, in a hydrogen-burning shell, Ne can be transmuted to Na when C is 
transmuted to N, even though the shell temperature is too low to permit 
O$\rightarrow$N conversion, and much too low to convert $^{24}$Mg to Al. 
The star L2208 is, however, anomalous if we require that all stars in 
M4 should begin their evolution with identical [el/Fe]-ratios. It is 
CN-strong (Table~\ref{Ivans.tab8}) and has the highest \carbiso\ ratio 
in our sample (although it still approaches the equilibrium value).  At 
the same time, it has relatively high abundances of O, Na, Mg and Al.  
We note that this one star defines the highest point in four of the elements 
shown in the boxplot (Figure~\ref{Ivans.fig7}) and that its abundances are 
in the upper quartile for yet four other elements.  Star L2208 also resides 
in a "special" place on the ($M_{bol}$,\teff) H-R diagram 
(Figure~\ref{Ivans.fig12}), assigned to neither the AGB nor RGB.  
Unfortunately, L2208 is among the stars we observed at slightly lower 
resolution (R $\sim$~30000), and, in this case, slightly lower 
signal-to-noise.  Nevertheless, it is possible that {\it this} 
star began with larger abundances of the $\alpha$- and light odd elements 
than its cluster companions and that it has undergone only rather modest 
deep-mixing, sufficient only to drive the \carbiso\ ratio toward the 
equilibrium value.  The existence of stars exhibiting primordial differences 
in the light element abundances would not be unique to M4.  For example, 
correlated CN and Na variations are seen in stars near the main sequence of 
47 Tuc (Briley 1997\markcite{Bri97}).  And, in M5, star IV-59 has a total 
C+N+O abundance exceeding that of any other analyzed giant in the cluster 
(Sneden \etal\ 1992\markcite{SKPL92}, Smith \etal\ 1997\markcite{SSBCB97}).

In Figure~\ref{Ivans.fig15}, we illustrate the correlation between 
Na and Al as well as the lack of one between Mg and Na (note the 
anomalous position of L2208, which we ignore for purposes of the 
subsequent discussion). If Na and Al are correlated, and if we assume 
that the Na enhancements are a result of proton captures on Ne, then 
one would suppose that the Al enhancements must be a result of proton 
captures on Mg, yet Mg seems unchanged. However, if we assume that the 
Al ``floor" for all M4 stars corresponds to those we observe having the 
lowest Al abundances ([Al/Fe] ~ +0.4, corresponding to 
{\rm log}~$\epsilon$(Al) $\sim$~5.7), then the Al enhancement required 
is +0.4~dex. The ``highest" Mg abundances we see is about [Mg/Fe] = 
+0.5 (corresponding to {\rm log}~$\epsilon$(Mg) $\sim$~6.9): the Mg 
abundance would need to drop by only about 0.05~dex to account for the 
increase in the abundance of Al. This change is too small to be detected 
with certainty from the present observational material. Such a change 
{\it may} be compatible with theoretical expectation, if scaled solar 
values for the isotopic ratios of $^{24}$Mg/$^{25}$Mg/$^{26}$Mg are 
invoked in the case of M4 (78.99:10.00:11.01, de Bi\`evre \& Barnes,
1985\markcite{dBB85}). 

As shown by Langer \& Hoffman (1995\markcite{LH95}, their Fig. 1), very 
modest hydrogen depletion of the envelope material ($<$ 3\%) 
can lead to an enhancement of Al by +0.4 dex when Na is enhanced by 
+0.7 dex, exactly as observed in our M4 sample. In this picture, the
enhancement of Al comes about entirely by destruction of $^{25}$Mg and 
$^{26}$Mg: $^{24}$Mg remains untouched.  Proton-capture on the lesser 
isotopes of Mg can occur at lower temperatures (T$_{9}~\sim~0.04$)
than on $^{24}$Mg (T$_{9}~\sim~0.07$). It is unclear whether the scaled 
solar values are appropriate in the case of the lesser isotopes of Mg.
There is evidence (Tomkin \& Lambert 1980\markcite{TL80}, Barbuy 
1985\markcite{Bar85}, Lambert \& McWilliam 1986\markcite{LM86}, Gay \&
Lambert 1999\markcite{GL99}) for a reduction in the abundances of the 
rare isotopes of Mg relative to $^{24}$Mg in $\sim$~20 to 30 metal-poor 
stars. On the other hand, Shetrone (1996b) found solar-like ratios of 
$^{25}$Mg+$^{26}$Mg to $^{24}$Mg among giants in M13, where the 
star-to-star abundance variations in Mg are found to be a result of 
processing $^{24}$Mg into Al.  We attempted to determine this in our
M4 stars by performing spectrum synthesis of the Mg isotope ratios in 
the MgH region near $\lambda$ 5140~{\rm AA}, adopting the line lists 
of Gay \& Lambert (1999\markcite{GL99}).  There is a redward asymmetry
of the MgH lines that seems to be better modelled by a solar or
slightly supersolar Mg isotopic ratio than a subsolar one.  While it
is tempting to interpret this as a similarity in the Mg isotopic 
ratio behaviour between the giant stars in M4 and M13 where the cluster
star behaviour is different from that found in the field, higher
signal-to-noise data and, more importantly, higher resolution data, 
perhaps of R $\sim$ 120,000, are required to determine the isotopic 
ratios of the M4 stars with complete confidence.

\subsubsection{C,N,O Abundances as a Function of Evolutionary State}

The Na {\it vs.} O and Na {\it vs.} Al correlations among the M4 giants 
permit us to comment on the possible evolutionary dependence of 
proton-capture element abundances to the C,N,O group. The range of 
bolometric magnitude over which we have a ``pure" sample of either RGB 
or AGB stars is not large ($\sim$~1~mag.), and the two branches cannot 
be distinguished when $M_{bol}$ $<$ --2.0 (see 
Figure~\ref{Ivans.fig12}). Consequently, we confine our attention 
to an RGB group of 11 stars, the brightest of which is L3612, and an 
AGB group also consisting of 11 stars, the brightest of which is L2206. 
Presumably, the AGB stars represent a stage of evolution more advanced 
than the RGB stars; we treat the groups as ``single" entities, without
any attempt at further subdivision. We designate L1408 and L2206 as 
``CN-weak" in accordance with their EW7874 values, even though previous 
investigations based on the blue CN-bands referred to them as 
``CN-strong".  In this exercise, we also deleted L2208, a star which we 
had already concluded has anomalously high primordial abundances of the 
$\alpha$- and light odd elements (see \S4.2.2).

We note here that the AGB population is over-represented in our sample 
relative to the RGB, since stellar evolutionary theory indicates that 
the AGB lifetime is only about 20 percent that of the RGB. This comes 
about entirely as a result of observational selection: in a 
spectroscopic survey, one tends to choose the brighter stars first. 
However, this should have no effect on the conclusions reached here, 
since the AGB and RGB samples are surely large enough to be 
representative of their evolutionary states. 

Although oxygen abundance estimates are available for all 22 of these 
stars, carbon and nitrogen abundances are available for only six of 
them.  However, all but four (all on the RGB) have CN-strength 
designations. We therefore explore the question of what combination of 
C,N,O abundances governs the CN-strong and CN-weak designations, on the 
basis of all ten stars for which C,N,O abundances are available 
(excluding L1514 and L2406 as discussed in \S3.3 and 4.1) in 
Table~\ref{Ivans.tab8}.  In Figure~\ref{Ivans.fig13}, all CN-strong 
stars have C depleted and N enhanced. If CNO processing occurs, O 
depletion will be accompanied by C depletion, which is also seen in 
Figure~\ref{Ivans.fig13}. And, the position of L4201 shows that it is 
possible to have a star in which O is not depleted, but still is 
CN-strong if C has been converted to N.  CN-weak stars seem to be 
driven almost exclusively by the low abundance of N, with C taking on 
almost any value. The low abundance of N, in turn, is controlled 
largely by the absence of significant O$\rightarrow$N conversion. The 
conclusion we reach is that CN-strong stars reflect overabundances of N, 
driven mostly by O$\rightarrow$N conversion, but occasionally by 
C$\rightarrow$N conversion only. One might therefore expect that 
CN-strong stars would become more frequent members of the ensemble if 
O$\rightarrow$N and C$\rightarrow$N conversions plus deep-mixing were 
coupled to advancing evolutionary state (as seems to have happened in 
M13 (Kraft \etal\ 1997\markcite{KSSSLP97}, Hanson \etal\ 
1998\markcite{HSKF98})).

This idea does not work for M4.  Defining an M4 star as ``oxygen-depleted" 
if [O/Fe] $<$ +0.20,\footnote{This choice of division between ``depleted" 
and ``undepleted" is meant to take into account the fact that the highest 
[O/Fe]-ratios are typically near +0.40, and the typical observational 
error is $\sim$~0.10 dex. Changing the division point by 0.05 to 0.10~dex 
would not change the conclusions reached here in a significant way.}
eight of the eleven RGB stars are undepleted whereas three are 
significantly depleted in oxygen. Of these, six are CN-strong, one is 
CN-weak, and four have unknown CN strengths. Most RGB stars therefore 
show little depletion of oxygen, yet a number are CN-strong, which 
suggests that they are similar to L4201, in which carbon has been 
significantly transmuted to nitrogen. On the other hand, of the eleven 
AGB stars in our sample, nine suffer little or no oxygen depletion and 
all are classified as CN-weak. Only two are both oxygen-depleted and 
CN-strong. Thus the AGB population in M4 shows much less evidence for 
deep-mixing than does the RGB population, even though the AGB is a later 
stage of stellar evolution.

What explanation can be offered for this apparent anomaly? Since we have 
not so tacitly assumed that the CNO abundances observed reflect a ``deep 
mixing" scenario, it is well to point out that this anomaly is not 
better explained by a ``primordial" scenario in which one supposes that 
all CNO (and by implication Ne, Na, Mg and Al) abundances were 
completely imprinted in the original material out of which the present 
low-mass cluster stars were made. Rather it is the deep mixing scenario 
that may offer the most hope.  Some time ago, Suntzeff 
(1981\markcite{Sun81}) showed that, in M3 and M13, the AGB is composed 
only of CN-poor stars which Suntzeff interpreted as evidence for 
significant mixing events prior to the evolution to the HB.  In another
study, Norris \etal\  (1981\markcite{NCFD81}) noted that there are few 
CN-strong AGB stars in NGC~6752, as well as a gap in the HB of that 
cluster.  They suggested that perhaps both effects not only had the same 
cause, but that the ``{\it ...high CN group does not ascend the giant 
branch for a second time}". Later theoretical and observational work tends 
to support this idea. Langer \& Hoffman (1995\markcite{LH95}) proposed 
that deep mixing would alter the He/H ratio in the envelope of a red giant 
and send the descendant onto the blue part of the HB. The basic idea was 
extended and put on a firmer basis in calculations by Sweigart 
(1997a\markcite{Swe97a},b\markcite{Swe97b}). A high-He star might fail 
to return to the AGB after occupation of the HB (e.g. HB evolutionary 
sequences of Dorman 1992\markcite{Dor92}).  In the case of M13, virtually 
all giants near the red giant tip show evidence of very deep mixing 
(Kraft \etal\ 1997\markcite{KSSSLP97}), and this in turn could be a major 
contributor not only to the ``blueness" of M13's HB but also to the 
relative paucity of AGB to RGB stars in that cluster (Caputo \etal\ 
1978\markcite{CCW78}, Buzzoni \etal\ 1983\markcite{BFBC83}). 

Does this scenario apply in the case of M4? Caputo \etal\ 
(1978\markcite{CCW78}) proposed that M4 has a low ratio of AGB to RGB 
stars just like M13; however, Buzzoni \etal\ (1983\markcite{BFBC83}) 
argued that the ratio is only slightly below average. Differential 
reddening across the face of M4 clouds the determination of this ratio 
and may account for the difference between the estimates. In addition, 
the depletion of O is much smaller in M4 than in M13, and, as we have 
asserted in \S4.2.2, the enhancement of Al in M4 is compatible with the 
destruction of $^{25,26}$Mg rather than the destruction of $^{24}$Mg. 
Thus if deep mixing is required in M4 to explain the change in the 
abundances of C, N, O, Na, Mg and Al with evolutionary state, then the 
depth of mixing required is less than that presumed to occur in M13. The 
calculations of Langer \& Hoffman (1995\markcite{LH95}) indeed suggest 
that the H envelope depletion needed in M4 is less than 3\%, 
whereas the requirement in the case of M13 appear to be closer to 15\% 
(Kraft \etal\ 1998\markcite{KSSSF98}). Thus it is unclear if ``mixed" 
M4 giants would ultimately evolve into HB stars that are too blue to 
return to the AGB.  Smith and Norris (1993\markcite{SN93}) observed seven 
RHB stars in M4 in a narrow {\it B-V} range.   Based on nitrogen-enhanced 
synthetic spectra modelling of the blue CN band, Smith \& Norris find very 
little CN variation and, with cautionary words about the assumed [O/Fe],
[C/Fe], and other uncertainties in the model-dependent result, suggest 
that the spectra of all seven stars might be analogs of the CN-strong red 
giants.  It would be of interest to observe these same seven stars using a 
larger wavelength coverage, to determine whether or not the stars still 
seem to be CN-strong in the red system, and to derive the abundances of 
the proton-capture nulceosynthesis elements, to determine whether the 
abundances of these much more evolved stars are consistent with our 
findings of the CN-strong/CN-weak abundance correlations among the giant 
branch stars.  We emphasize once again that a purely primordial scenario 
would fare even worse in explaining the evolutionary abundance changes: 
only one of seven (relatively unevolved) RGB stars in our sample are 
designated as CN-weak.

\section{Comparisons with Some Other Clusters}
\subsection{The Na {\it vs.} O Anticorrelation}
\subsubsection{M4 {\it vs.} M5}

We first compare the Na {\it vs.} O anticorrelation of M4 with that of
M5, a cluster having essentially the same metallicity as M4, and which 
was previously analyzed using techniques (Sneden \etal\ 
1992\markcite{SKPL92}) similar to those employed here.  In 
Figure~\ref{Ivans.fig16}, we plot [Na/Fe] {\it vs.} [O/Fe] for the 
giants in these two clusters. Recall that we believe that, (a) L2208
is the one M4 giant in which the abundances of the $\alpha$- and light 
odd elements are enhanced ab initio, and (b) L4201 is also the only M4 
giant in which the CN-strong phenomenon is solely a result of 
C$\rightarrow$N conversion, O playing essentially no role. Eliminating 
these two stars (as well as the suspected binary L2406), we find that 
each cluster exhibits a noticeable Na {\it vs.} O anticorrelation, but 
the relationships are somewhat offset.  The least ``mixed" stars of M4 
have higher O and higher Na abundances than those in M5 (by $\sim$~0.1 
and $\sim$~0.2~dex, respectively), and the gap widens in the most mixed 
stars (to $\sim$~0.3 dex and $\sim$~0.25 dex, respectively). We note 
that, at comparable luminosities, M5 stars tend to have higher derived 
\teff\ values than M4 stars.  If there were an error in the temperature 
determinations for one cluster relative to another, could this generate
the offsets illustrated in Figure~\ref{Ivans.fig16}? No, because from 
the \teff\ dependencies listed in Table~\ref{Ivans.tab6}, a change in 
\teff\ scales to force agreement between M4 and M5 would produce 
abundance changes having a direction opposite to that required. We 
therefore conclude that the offsets are probably real, and therefore 
that the ab initio O and Na abundances of M4 are a bit higher than those 
of M5. It has been amply demonstrated elsewhere ({\it e.g.} Pilachowski 
\etal\ 1996\markcite{PSK96}, Kraft \etal\ 1997\markcite{KSSSLP97}) that 
there is a range of Na and O abundances among halo field giants which 
show little or no evidence of deep-mixing.  It should, therefore, come 
as no surprise that the least mixed stars of globular clusters might 
exhibit variations in the abundances of these elements on the average 
from cluster to cluster.

A point of similarity exhibited in Figure~\ref{Ivans.fig16} is that the 
{\it range} of the O and Na variations is about the same in the two 
clusters: for O, a factor of $\sim$~2.5 in M4 and a factor of $\sim$~3 
in M5. Correspondingly, the Na factors are $\sim$~4.5 and $\sim$~4, 
respectively. According to the deep mixing models of Langer \& Hoffman 
(1995\markcite{LH95}, their Fig. 1), a depletion of O by factors of 2.5 
and 3 leads to enhancements of Na by factors of 4 and 5, respectively, 
in excellent agreement with what is observed.  The models assume, of 
course, that Ne, in addition to the other $\alpha$-elements, is enhanced 
by 0.4~dex in [Ne/Fe] above the solar value. We warn the reader that 
there is no way to determine spectroscopically if [Ne/Fe] = +0.4 or if 
the Ne abundance is the same in M4 as in M5, even if such assumptions 
are plausible. Since we argue that the ab initio O and Na might be 
slightly different in the two clusters, there is no assurance that Ne is
not also slightly different. 

In Figure~\ref{Ivans.fig17}, we exhibit the Na {\it vs.} O 
anticorrelation found in a number of intermediately metal-poor clusters 
having [Fe/H] near --1.6 (see the figure caption for references). We 
omit M5, but superimpose the results from the present study of M4. 
Generally, we see that M4 is slightly offset from the other clusters: 
its ab initio Na and O abundances are a bit higher than in the other 
clusters, echoing its relationship to M5, although the differences are 
less extreme.  The behavior of NGC 7006 seems closest to that of M4.

\subsection{The Mg {\it vs.} Al Anticorrelation}

In Figure~\ref{Ivans.fig18}, we plot [Al/Fe] {\it vs.} [Na/Fe] 
for our M4 sample as in Figure~\ref{Ivans.fig15}, but we now add M5 
(Shetrone 1996a\markcite{Sh96a}) and the two slightly more metal-poor 
clusters M13 (Kraft \etal\ 1997\markcite{KSSSLP9}) and NGC~7006 (Kraft 
\etal\ 1998\markcite{KSSSF98}).

In M4, the range of variations of Al ($\sim$~0.4~dex) and of Na 
($\sim$~0.6~dex) are internally consistent with the deep-mixing 
calculations of Langer \& Hoffman (1995\markcite{LH95}, their Fig. 1), 
which assume that the enhancements of Al and Na are generated at the 
expense of $^{25,26}$Mg and $^{22}$Ne, respectively. The NGC~7006 
sample is quite small (six stars), but the range of both Al and Na is 
similar to that of M4, as is the ``floor'' of both [Na/Fe] and [Al/Fe].  
This suggests that the primordial sources of Al and Na (as well as 
their subsequent enhancements) in these two clusters may have a 
similar history. The distributions in M5 and M13 are, however, quite 
different from those of M4 and NGC 7006 where the abundance ``floor" 
is clearly much lower for both [Na/Fe] and [Al/Fe] ($\sim$~--0.25~dex 
and $\sim$~--0.1~dex, respectively). The range for M13 is very large 
and it has been argued elsewhere (Shetrone 
1996a\markcite{Sh96a},b\markcite{Sh96b}, Pilachowski 
\etal\ 1996\markcite{PSK96}, Kraft \etal\ 1998\markcite{KSSSF98}) that 
such a large range could only be induced if deep-mixing brought up the 
product of proton captures on $^{24}$Mg. Although the ranges of Al and 
Na in M5 are much smaller than in M13, the current sample size in M5 is 
too small to claim that $^{24}$Mg-depletion, as in M13, is ruled out.  
Clearly a larger sample of M5 giants is needed before a definitive 
picture can be established.

We plot [Al/Fe] {\it vs.} [Mg/Fe] in Figure~\ref{Ivans.fig19}, 
once again superimposing data from M5, M13, M15 and NGC 7006 from the 
references noted above. As in Figure~\ref{Ivans.fig18}, the 
range and location of the NGC 7006 data is similar to that of M4. In 
contrast to M4, M13 shows the large range of Al abundances compatible 
with destruction of $^{24}$Mg (Shetrone 1996b\markcite{She96b}). 
Striking, however, is the noticeable offset of Mg abundances in M5 
with respect to M4, a reduction of Mg abundances by a factor of 2, on 
the average. Of the six M5 stars in the Shetrone sample, two are O-rich 
and four are O-poor (using the definition of \S4.2.3); the CN 
classification is known for only two of these (Smith \etal\ 
1997\markcite{SSBCB97}, Smith \& Norris 1983\markcite{SN83}). Within 
the errors, the Mg abundances show little or no spread between the O-rich 
and O-poor subgroups. Once again, this suggests that among these stars, 
there has been little or no destruction of $^{24}$Mg, independent of 
destruction of O. This is turn suggests that the {\it primordial 
abundance of Mg} (dominated by $^{24}$Mg) {\it is higher in M4 than in 
M5}, despite the fact that the two clusters have virtually the same iron 
abundance.  Again, this may not be surprising: among field halo giants 
(Pilachowski \etal\ 1996\markcite{PSK96}, Hanson \etal\ 
1998\markcite{HSKF98}), stars which exhibit no evidence for conversion 
of Ne to Na or Mg to Al, and therefore presumably reflect their original 
primordial abundances of these elements, the Mg and Na abundances are 
correlated. This indicates that there is a substantial primordial range 
in both of these elements at a given metallicity among halo field stars. 
Therefore it should come as no surprise that differences in the ab 
initio abundances of [Mg/Fe] as well as [Na/Fe] should exist from one 
cluster to another.

\subsection{Alpha-Element Variations}

To the evidence that Mg abundances in M4 exceed that in M5 by a factor
of 2, we add the evidence that {\it silicon abundances in M4 also 
exceed that of M5 by a factor of 2}: $<$[Si/Fe]$>_{M4}$ = +0.55 
$\pm$~0.02 (Table 5, 23 stars) {\it vs.} $<$[Si/Fe]$>_{M5}$ = +0.20 
$\pm$~0.02 (Sneden \etal\ 1992\markcite{SKPL92}; 13 stars). Ca and Ti 
abundances in the two clusters are essentially the same and have the 
``usual" modest overabundances (relative to Fe) in the +0.2 to +0.3 
range.  The $\alpha$-elements of M4 are therefore somewhat unusual 
among clusters in the intermediately metal-poor range but mimic those 
found in the very metal-poor cluster M15 for which [Mg/Fe] $\sim$~+0.6 
(Sneden \etal\ 1997\markcite{SKSSLP97}; for the 12 stars that have not 
experienced Mg depletion), $<$[Si/Fe]$>$ = +0.60 (12 stars), but 
$<$[Ca/Fe]$>$ = +0.24 (18 stars). In M15, $<$[Ti/Fe]$>$ = +0.46 $\pm$ 
0.12 but the result is not very reliable since it is based on only 
three stars.  This Mg/Si ``overabundance" {\it vs.} Ca/Ti ``normal" 
abundance anomaly of M4 {\it vs.} M5, is seemingly accompanied by a 
correspondingly high ``floor" of Al abundances, also found in M15. 
Substructure in $\alpha$- and light odd elements is found also among 
relatively metal-rich ([Fe/H] $>$ --1) disk dwarfs (\eg, Edvardsson 
\etal\ 1993\markcite{EAGLNT93}; {\it cf.}, Tomkin \etal\ 
1997\markcite{TELG97}) and galactic nuclear bulge giants with 
metallicities near [Fe/H]~=~--1 (McWilliam \& Rich 1994\markcite{MR94}), 
but the distribution of the substructure does not always match that of 
either M4 or M5.  For example, the galactic bulge giant BW IV-003 
([Fe/H]~=~--0.94) mimics the distribution of the $\alpha$-elements and 
Al in M5, whereas BW IV-329 ([Fe/H]~=~--0.85) is similar to M4. However, 
there are other bulge giants in which these [el/Fe]-ratios are 
dissimilar to both M4 and M5. And, in the disk dwarf sample of 
Edvardsson \etal\markcite{EAGLNT93}, Ca and Si are paired as rising on 
the average more slowly with decreasing [Fe/H] than Mg and Ti.

What does seem clear is that the differences between Mg, Si, Al and Na
in M4 and M5 arise from some property of the primordial nucleosynthetic
sites. Mg, Na and Al are the products of Ne- and C-burning. The 
$\alpha$-elements are rather insensitive to the initial metallicity of 
the massive progenitor stars, but the abundances of both Na and Al are 
controlled by the neutron flux during the Ne- and C-burning. One 
expects [Al/Mg]~=~$C_{1}$[Mg/H], and [Na/Mg]~=~$C_{2}$[Mg/H], where 
$C_{1}$ and $C_{2}$ are constants $>$~0 (Edvardsson \etal\ 
1993\markcite{EAGLNT93}). Considering the least ``proton-capture 
enhanced" stars in both M4 and M5, we derive $C_{1}$~=~0.15 for M5 and 
0.14 for M4.  That is, the [Al/Fe] ratios for these two clusters are 
internally consistent if it is assumed that the Mg and Al abundances 
arise from a primordial nucleosynthetic site as described above. In the 
case of Na, we are less successful: on the same assumptions, we find 
$C_{1}$~=~0.45 for M5 but $C_{2}$~=~1.00 for M4. However, the least 
``enhanced" floor for Na is rather poorly known for M5, so the 
difference between the two values of $C_{2}$ may not be too 
significant.

\subsection{Ba, La, and Eu Abundances in M4, M5 and $\omega$ Cen}

Among halo field giants, and giants in clusters such as M71, M13 and
M92, $<$[Eu/Fe]$>$ ranges from +0.3 to +0.5 with little scatter within 
a given cluster (Shetrone 1996a\markcite{Sh96a}). For M5, Shetrone 
found $<$[Eu/Fe]$>$ = +0.44 $\pm$~0.03, based on six stars. Thus the 
value we derive in the case of M4 (Table~\ref{Ivans.tab5}), 
$<$[Eu/Fe]$>$ = +0.35 $\pm$~0.02 is in no way unusual. What is 
surprising in M4 is the high abundance of Ba: $<$[Ba/Fe]$>$ = +0.60 
$\pm$~0.02 (Table~\ref{Ivans.tab5}), which is supported by the high
abundance of La: $<$[La/Fe]$>$ = +0.45 $\pm$~0.02.  The [Ba/Eu] ratio, 
often taken as a measure of $s$- to $r$-process nucleosynthesis in the 
primordial material of the cluster, has the unusually high value of 
+0.25. Generally, field halo and globular cluster (M5, M13, M92) 
giants agree (Gratton \& Sneden 1991\markcite{GS91}, 
1994\markcite{GS94}, Armosky \etal\ 1994\markcite{ASLK94}, Shetrone 
1996a\markcite{Sh96a}) that [Ba/Eu] is typically negative with a range 
from --0.2 to --0.6.  The [Ba/Eu] ratio in M4 is more than four times 
higher than that of the ``normal" cluster M5, and is 0.25~dex higher 
than the total solar-system {\it r~+~s} value.  

We performed numerical experiments by combining our derived Ba, La, 
and Eu abundances using a solar-system $r$-process ratio among the 
elements (Cowan, 1999\markcite{Cow99}).  We found that the $s$-process
ratios of our individual stars agree fairly well to within the abundance 
errors and are enhanced over the solar ratios for these elements.  The Ba 
abundance in M4 cannot be attributed to the $r$-process component; in the 
M4 stars, we have a larger $s$:$r$-process contribution than in the sun.  
Furthermore, there is no dependence of [Ba/Fe] on evolutionary state in 
M4, \ie, [Ba/Fe] on the AGB does not exceed [Ba/Fe] on the RGB. Therefore 
the Ba-excess cannot result from neutron captures on Fe-peak elements 
during a He shell flash episode on the AGB, but must be a signature of 
$s$-process enrichment of the primordial material out of which the 
low-mass M4 stars were formed.  As emphasized by the referee, the 
presence, and especially the excess, of the $s$-process elements provides 
evidence that the period of star formation and mass-loss that preceded 
the formation of the observed stars in M4 was long enough for AGB stars 
to contribute their ejecta into the ISM of the cluster. This, in turn, 
implies that stars of intermediate mass, of 3--10 solar masses, had time 
to evolve.  Contribution from the $s$-process is very well evidenced in 
the globular cluster $\omega$ Cen (Vanture \etal\markcite{VWB94}).

It is also the case that in the multimetallicity cluster $\omega$ Cen, 
there are many giants with Ba abundances that are $\sim$~0.6~dex 
higher than in clusters such as M5, M13 and M92 (Norris \& Da Costa 
1995b\markcite{ND95b}).  At the same time, Smith \etal\ 
(1995\markcite{SSBCB97}) demonstrated that some $\omega$ Cen giants 
also had unusually low [Eu/Fe]-ratios (between --0.2 and --0.7), so 
that the high [Ba/Eu] values were in some instances a result of a 
deficiency in Eu rather than an excess of Ba.  Clearly $\omega$~Cen 
shares some abundance characteristics in common with M4 (there exists
in $\omega$ Cen a subset of stars which have nearly identical elemental 
overabundances with respect to iron as seen in M4 for aluminum, silicon, 
barium, and lanthanum), but it also possesses a more complicated 
nucleosynthetic history than M4. The important point here is that the 
high Ba and La properties of M4 stars is surely a primordial, not an 
evolutionary, effect.

\section{Summary}

We have conducted a large-sample high and medium resolution 
spectroscopic survey of 35 giant stars in the nearby mildly 
metal-poor ([Fe/H] = --1.18) globular cluster M4. In studying M4, we 
were confronted with a cluster having interstellar extinction that is 
large and variable across the cluster face, and which may obey a 
non-standard extinction law.  Therefore, we combined traditional 
spectroscopic analytical techniques with new approaches to temperature 
determinations in order to derive a self-consistent set of atmospheric 
parameters and abundances for our program stars.  Using our spectroscopic 
results, we derive a total-to-selective extinction ratio of 3.4~$\pm$~0.4 
as well as an average $<$E(B--V)$>$ reddening significantly lower than 
that estimated by using the dust maps made by Schlegel \etal\ 
(1998\markcite{SFD98}).  Some caution should be used when using the dust 
maps to estimate reddening and extinction in regions of high extinction.

The derived abundances are summarized at the beginning of \S4, and 
will not be repeated in detail here. We find evidence that silicon 
and aluminum are primordially overabundant by factors greatly 
exceeding ($\sim$~2 to 3) the mild overabundances (relative to iron) 
usually seen in $\alpha$- and light odd elements among halo field and 
globular cluster giants of comparable metallicity, such as M5 and M13. 
We also find that barium is overabundant by a factor of about 4, and 
show that the [Ba/Fe]-ratio is independent of evolutionary state among 
M4 giants, thus ruling out that the excess Ba is a result of nuclear 
processes occurring within the M4 giants themselves.  These 
overabundances confirm the results of an earlier study (Brown \& 
Wallerstein 1992\markcite{BW92}) based on a much smaller sample of M4 
giants.

Superimposed on the somewhat unusual primordial abundance distribution
nevertheless is evidence for the existence of proton capture synthesis
of carbon, oxygen, neon, and magnesium, which leads to the production 
of nitrogen from C and O, sodium from Ne and Al from Mg, probably as a 
result of deep-mixing of the stellar envelope through the 
hydrogen-burning shell.  We recover the Na {\it vs.} O anticorrelation 
and Na {\it vs.} Al correlation found in other globular clusters.  The 
the range of variation is muted as compared with more metal-poor 
clusters such as M13 and M15 but larger than that found in the more
metal-rich cluster, M71.  The rather small range of variation of Al in 
M4, plus the absence of any significant variation in Mg, is compatible 
with the idea that the Al enhancements are derived from the destruction 
of $^{25,26}$Mg, in contrast to the destruction of $^{24}$Mg required 
in the case of M13.  While the abundance swings of C, O, Na, Mg and Al 
appear to be muted compared with what is seen in more metal-poor 
clusters, they have not entirely disappeared.

A puzzle arises from the fact that giants of the AGB have C,N,O 
abundances that show less evidence for proton capture nucleosynthesis 
than is found in the less-evolved stars of the RGB. We discuss the 
idea that deeply mixed stars of the RGB, subsequent to the He core 
flash, might take up residence on the blue end of the HB, and thus 
fail to evolve back to the AGB. Reasons for skepticism concerning this 
scenario are noted. On the other hand, attribution of this puzzle to a 
primordial scenario would fare even worse.

\section{Acknowledgements}

We are indebted to Jerry Lodriguss for generously sharing the electronic 
files containing his excellent deep sky images of the M4 Sco-Oph region.  
We also gratefully acknowledge Jeff Brown for supplying his EWs of three 
M4 stars; Tim Davidge and Flavio Fusi Pecci \& Francesco Ferraro for 
sharing their photometry of M4; and Pamela Gay \& David Lambert for 
notifying us of the results of their MgH work prior to publication as 
well as making their line lists available.  David Schlegel has our 
appreciation for answering questions regarding the dust map code as does 
David Gray for providing helpful feedback regarding the applicability of 
the spectroscopic line ratio technique.  We thank David Lambert, John 
Norris, Jim Truran, Neal Evans, John Lacy, Don Winget, Craig Wheeler,
Jos Tomkin, and Vincent Woolf for helpful discussions and/or thoughtful 
comments on drafts of this paper. We also appreciate the valuable 
suggestions and remarks made by the referee, George Wallerstein, and the 
subsequent improvements that these also made to the paper, in particular, 
the recommendation to determine and analyze the La abundance in the 
context of the Ba enhancements.

IRAS fluxes were obtained using IBIS, an observational planning tool for 
the infrared sky developed at IPAC.  The Infrared Processing and Analysis 
Center is operated by the California Institute of Technology and the Jet 
Propulsion Laboratory under contract to NASA.  This project has also made
use of NASA's Astrophysics Data System Abstract Service.  This research 
was made possible by NSF grants AST-9217970 to RPK, AST-9618364 to CS, and 
AST-9618459 to VVS.

Ed Langer, our co-author, passed away on Feb. 16, 1999, following a brief 
illness.  Ed's kindness and generous spirit added a great deal to our 
collaboration; his way of thinking about things clarified the basic 
physics at the heart of our observations and added a great deal to the 
science.  He will be greatly missed.

\clearpage

\begin{table}
\dummytable\label{Ivans.tab1}
\end{table}

\begin{table}
\dummytable\label{Ivans.tab2}
\end{table}

\begin{table}
\dummytable\label{Ivans.tab3}
\end{table}

\begin{table}
\dummytable\label{Ivans.tab4}
\end{table}

\begin{table}
\dummytable\label{Ivans.tab5}
\end{table}

\begin{table}
\dummytable\label{Ivans.tab6}
\end{table}

\begin{table}
\dummytable\label{Ivans.tab7}
\end{table}

\begin{table}
\dummytable\label{Ivans.tab8}
\end{table}

\begin{table}
\dummytable\label{Ivans.tab9}
\end{table}

\begin{center}
{\bf Figure Captions}
\end{center}
 
\figcaption[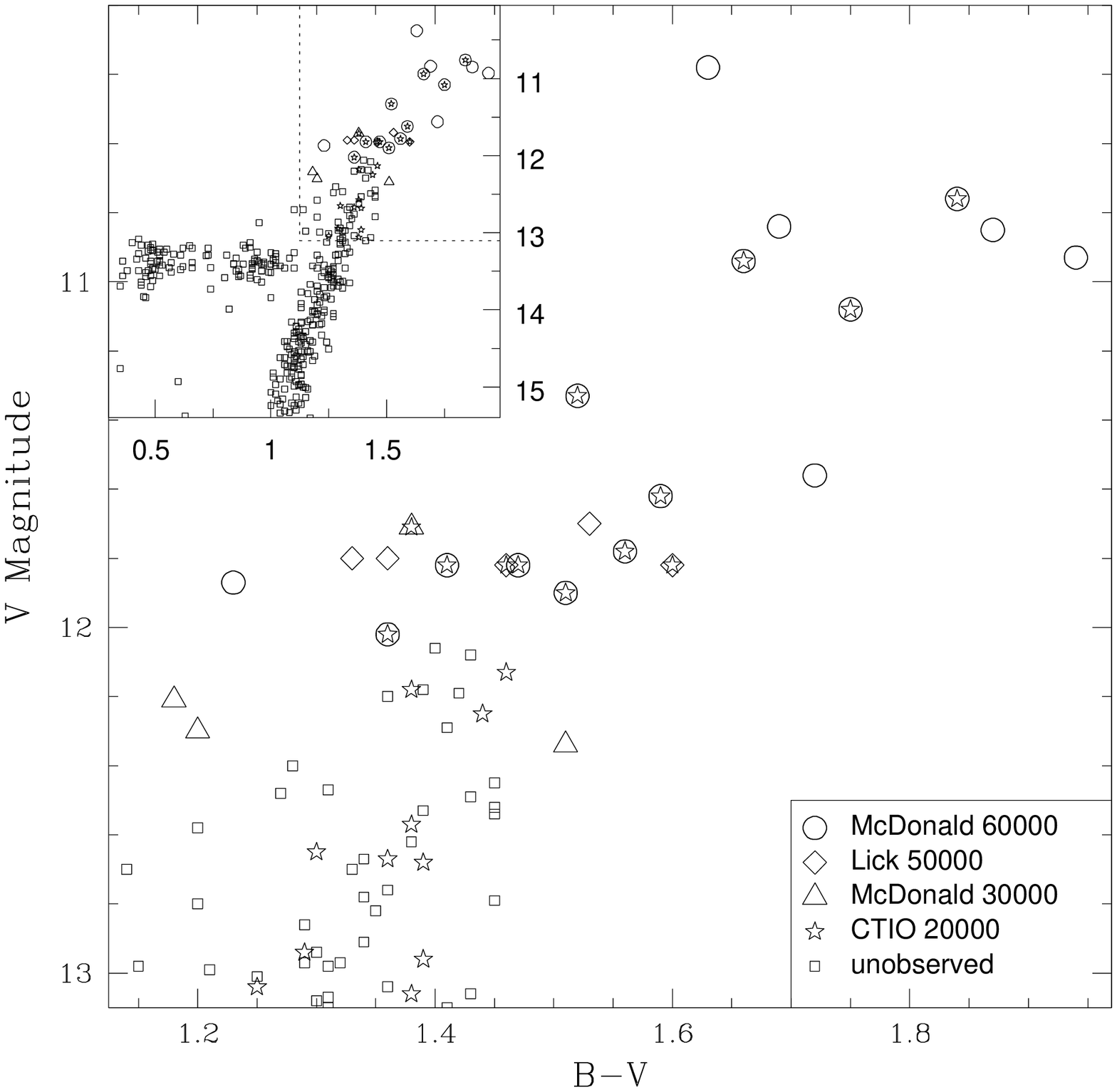]{A color-magnitude diagram of M4, with photometry 
from Cudworth \& Rees (1990), showing the positions of our program stars
on the giant branch.  The symbols are given in the figure legend and 
correspond to the observatory and resolution of the spectrograph used for 
each observation.  The inset diagram shows the program stars plotted in 
relation to all Cudworth \& Rees M4 stars of magnitude $\leq$~15.5.
\label{Ivans.fig1}}

\figcaption[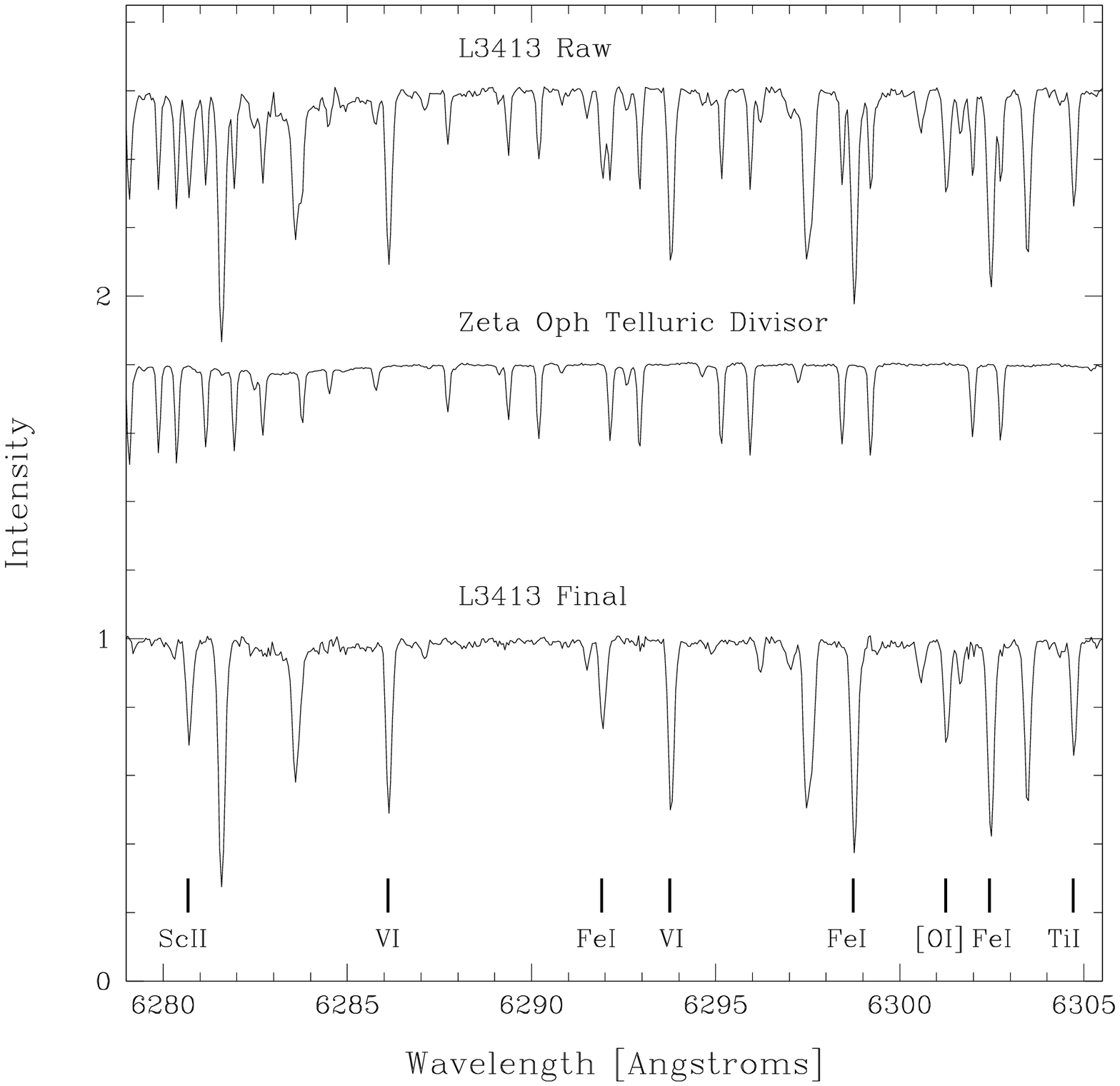]{A reduced, ``raw'', normalized spectrum of M4 
L3413 in the region near the important \wave{6300.3} \nion{O}{i} line, a 
``telluric standard'' hot, rapidly rotating divisor star (in this case
$\zeta$~Oph), and the ``final'' quotient spectrum, essentially free of 
telluric lines.  The relative flux scale of the L3413 is correct, and the 
other spectra have been shifted vertically by additive constants for 
display.
\label{Ivans.fig2}}

\figcaption[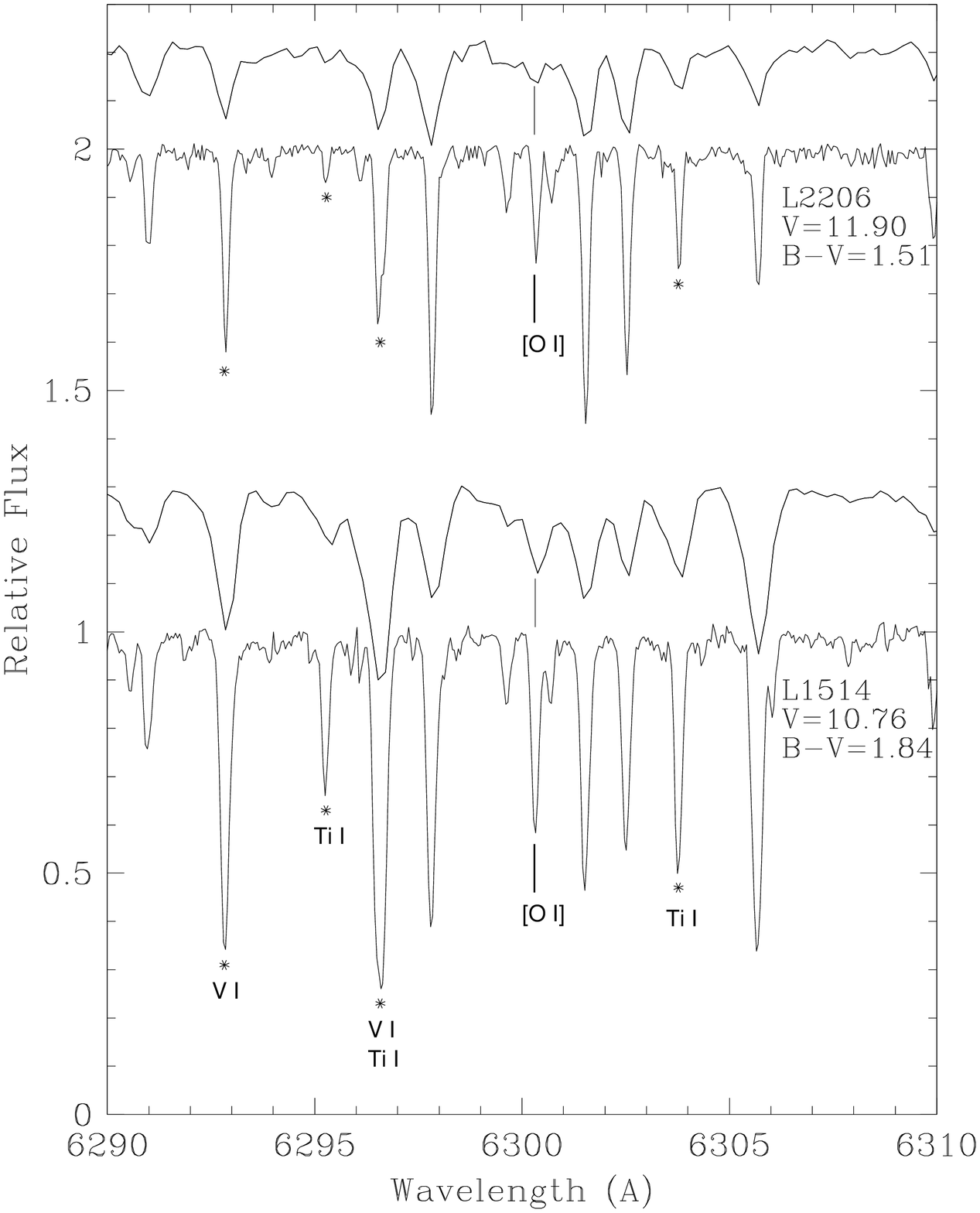] {High resolution 2d-coud\'e and medium 
resolution Argus spectra surrounding the \wave{6300.3} [\nion{O}{i}] line 
in M4 stars L1514 (an RGB tip star) and L2206 (a star near the 
low-luminosity limit of the high resolution data).  The relative flux 
scale of the L1514 high resolution spectrum is correct, and the other 
spectra have been shifted vertically by additive constants for display.  
The \nion{O}{i} line and several temperature-sensitive \nion{Ti}{i} and 
\nion{V}{i} lines are marked in the figure.  Most of the remaining 
prominent absorption features are due to \nion{Fe}{i}.
\label{Ivans.fig3}}

\figcaption[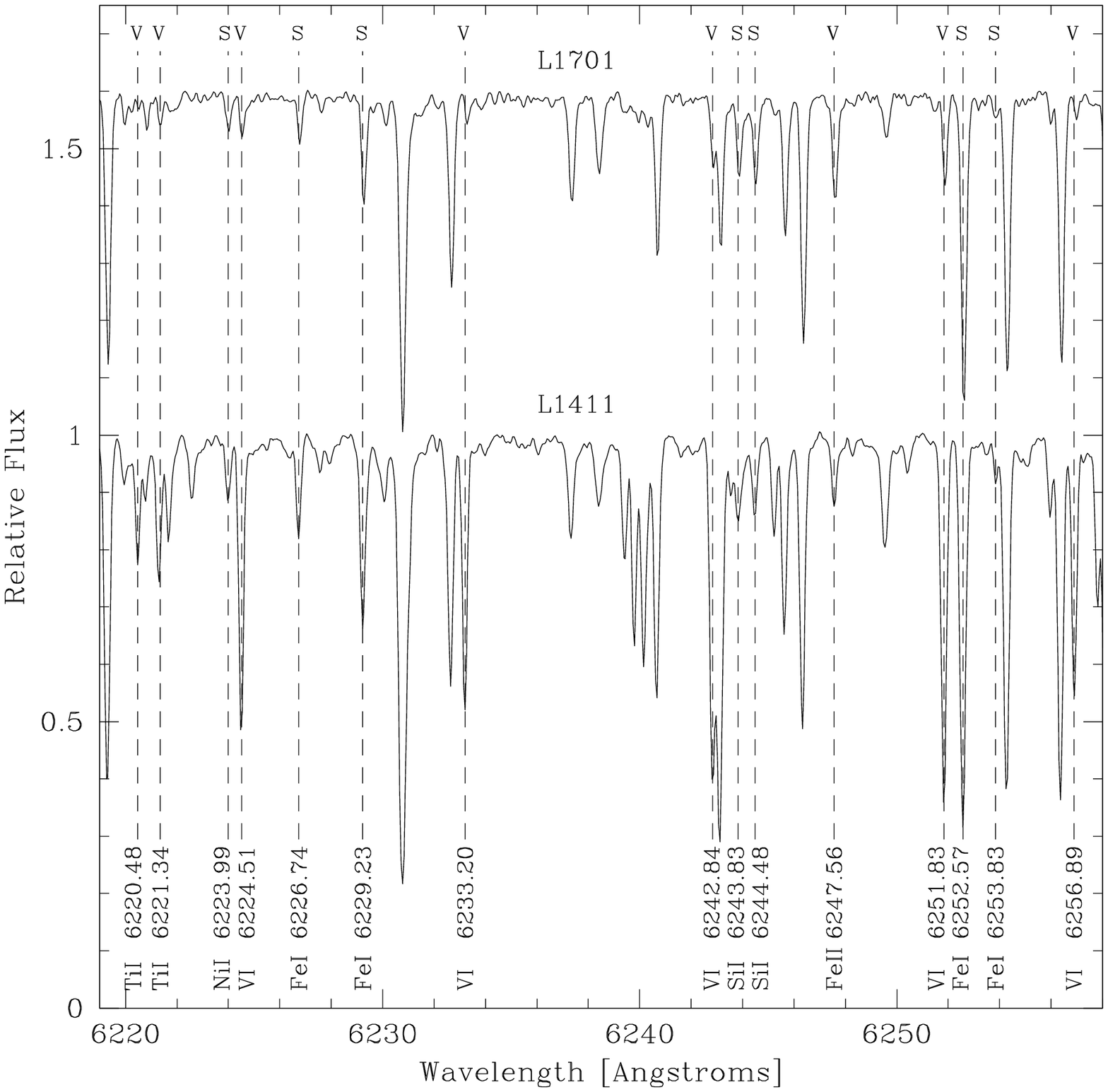] {Spectroscopic lines whose ratios are
good \teff\ indicators: M4 stars L1701 (4625K) and L1411 (3950K)
are shown, with a vertical offset added to the L1701 spectrum.
Lines that are temperature-variable according to Gray (1994) are 
indicated with ``V'' labels, and those that are temperature-stable
are marked with ``S'' labels.
\label{Ivans.fig4}}

\figcaption[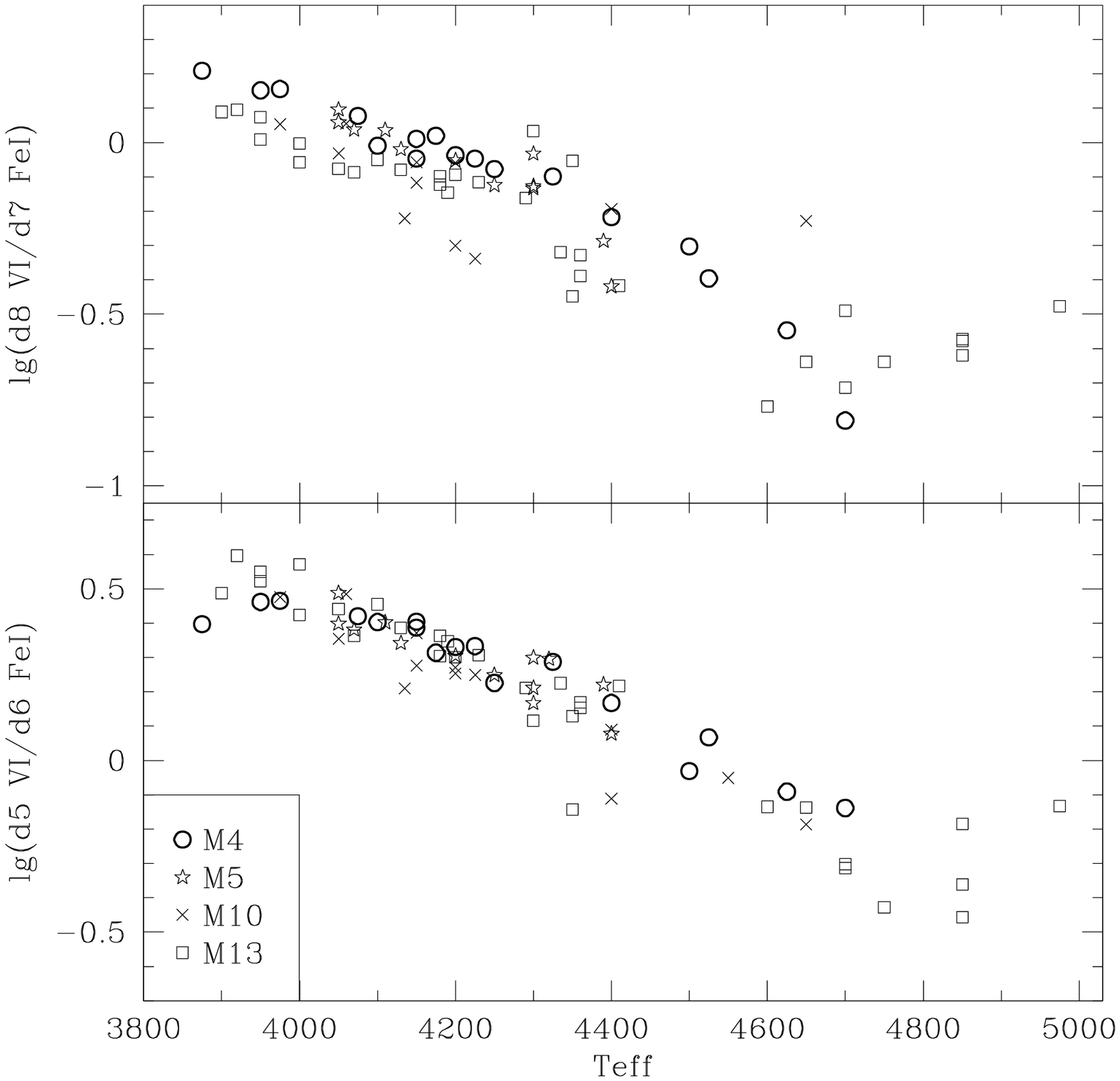] {Two typical line-depth ratios plotted as
functions of \teff\ in M4 and three comparison clusters observed
in earlier papers of this series (a comprehensive investigation of 
the spectroscopic line ratios of these data sets is underway and will be 
reported in a future publication).  The ratios are presented in logarithmic 
form, and the dividend and divisor lines are labeled both by Gray's (1994) 
numbering scheme (\eg, d5, d6) and by atomic species.  The scatter in the 
relationships with respect to \teff\ increases at higher temperatures 
because the line depths (especially those of the dividend lines) become 
weak just as the typical S/N of the spectra are becoming less in these 
fainter stars. 
\label{Ivans.fig5}}

\figcaption[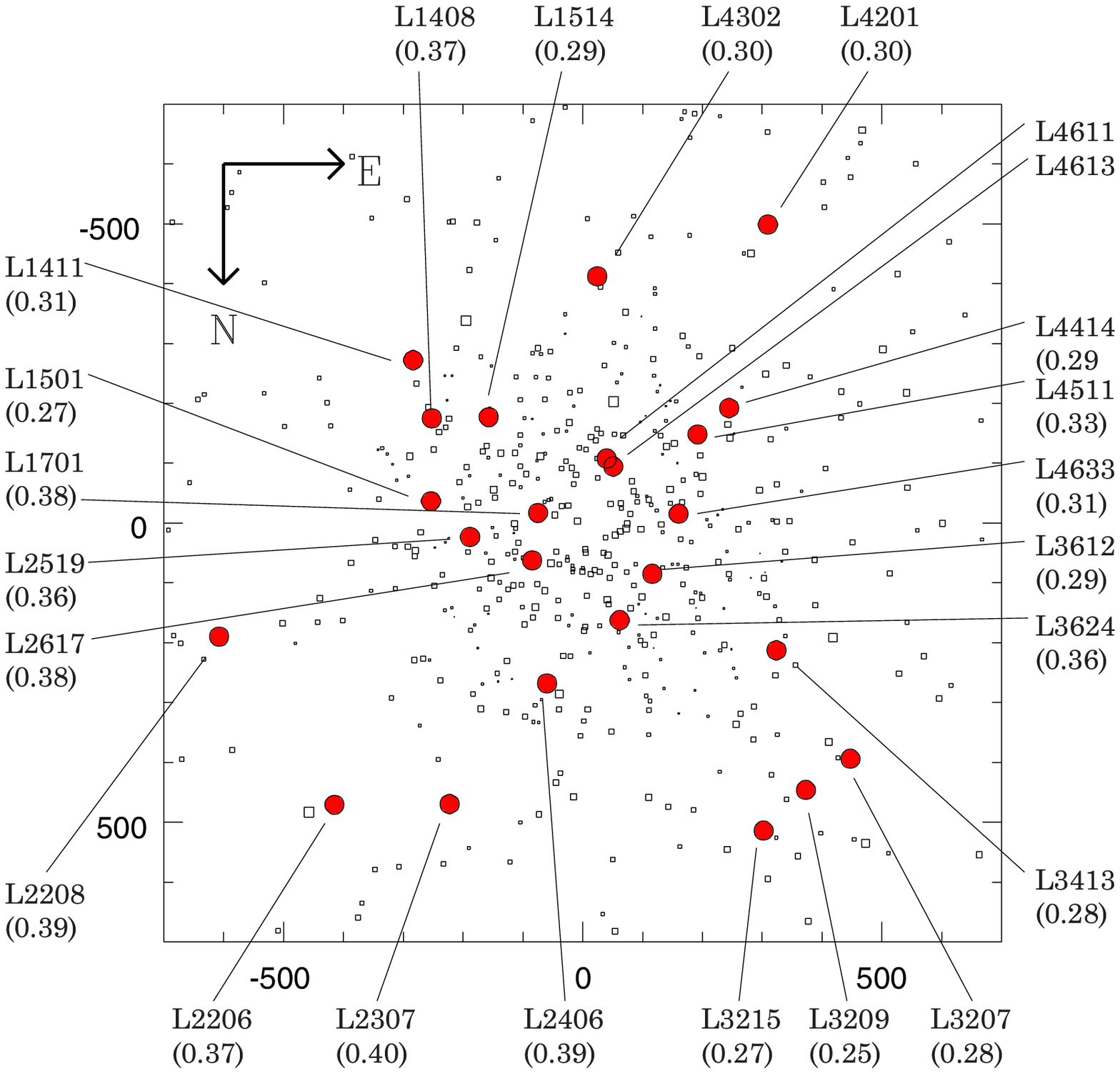] {A map of M4, showing the positions of the program 
stars we observed at high resolution using the Lee (1977) 
stellar identification scheme.  The relative stellar position information 
in seconds of arc is taken from Cudworth \& Rees (1990).  
Our 2d-coud\'e observations are denoted by shaded circles.  The remaining 
stars are denoted by hollow squares (where the size of the square roughly 
indicates a star's relative magnitude).  We also show our derived $E(B-V)$ 
estimates.
\label{Ivans.fig6}}

\figcaption[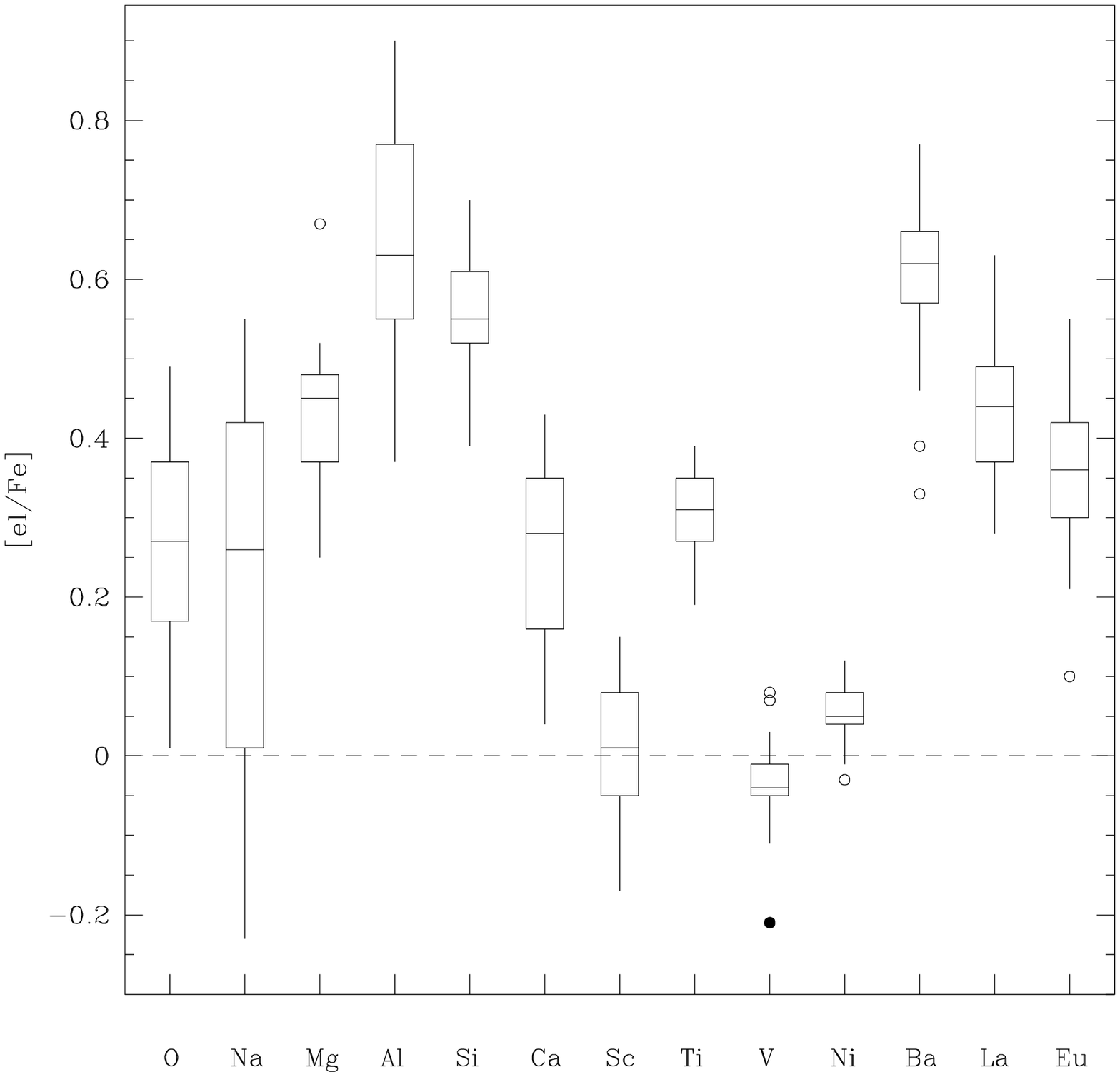] {A boxplot of the M4 giant star element 
abundances.  A boxed horizontal line indicates the interquartile range 
(the middle 50\%
of the data) and median found for a particular element.  The vertical 
tails extending from the boxes indicate the total range of abundances 
determined for each element, excluding outliers.  Mild outliers (those 
between 1.5$\times$ and 3$\times$ the interquartile range) are denoted 
by hollow circles (o) and severe outliers (those greater than 3$\times$ 
the interquartile range) by filled circles ($\bullet$).  The dashed line 
at [el/Fe] represents the solar value for a particular elemental 
abundance ratio.
\label{Ivans.fig7}}

\figcaption[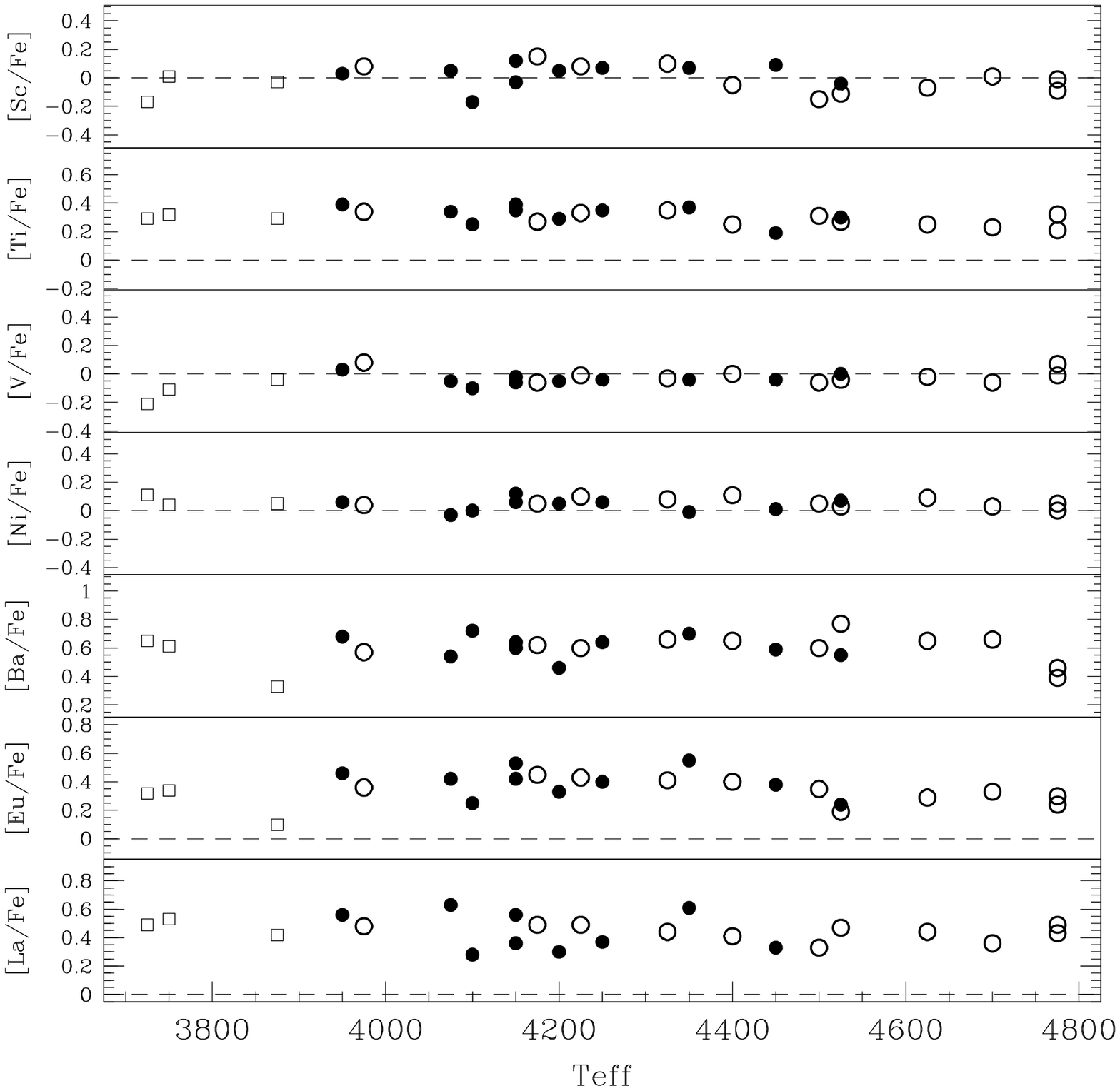]{Abundances of the Fe-peak and neutron-capture 
elements (Sc, Ti, V, Ni, Ba, La, and Eu) with respect to the iron abundance 
as functions of effective temperature.  A dashed line at [el/Fe]~=~0.0 
represents the solar value for a particular elemental abundance.  CN-weak 
stars are denoted by open circle symbols (o) and CN-strong stars are 
denoted by filled circles ($\bullet$).  Hollow squares ($\Box$) denote the 
values derived for the three program stars cooler than 3900~K, for which 
our abundances are not considered to be as reliable.
\label{Ivans.fig8}}

\figcaption[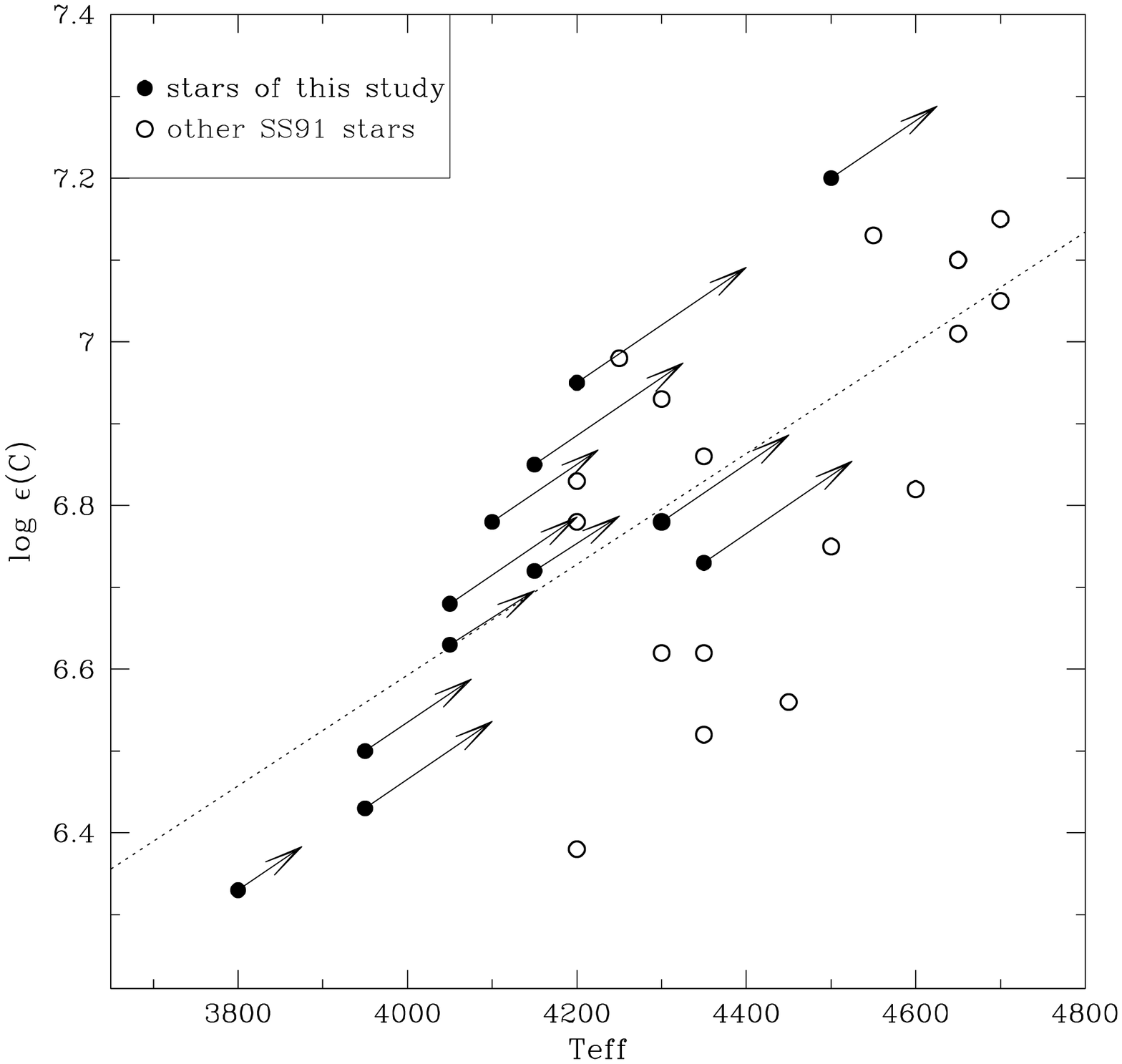]{Carbon abundances derived by SS91 plotted
versus their \teff\ values.  For the stars observed in our study, denoted
by filled circles ($\bullet$), we indicate with arrows the adjusted C 
abundances for the indicated higher \teff's of this work.
\label{Ivans.fig9}}

\figcaption[Ivans.fig10]{Representative spectra of the CN (2--0) 
bandhead region at 7874~{\rm \AA} of four of our M4 stars and a
``telluric standard'' star, $\zeta$ Oph.  The spectra between the two
vertical dashed lines illustrate the region we used to measure EW.
The stars presented are listed in the following order, from top to bottom: 
L4511, a star which appears to be very CN-strong in both the red and the 
blue measures; L3413, a star of similar \teff\ but of very weak CN-strength 
(as measured in the red); L4414 and L4633, a warmer pair of stars, showing 
a similar contrast in CN-strength at a higher \teff.
\label{Ivans.fig10}}

\figcaption[Ivans.fig11] {A comparison of the S(3839) index from
Norris (1981) or SS91 (or a mean for the stars in common) and our EW7874
measure, plotted as a function of the observed magnitude.
The solid circles ($\bullet$) in the upper plot are those stars in common 
with this study.  The lower plot includes all of the stars observed at 
high resolution used in this study.  The baseline for the S(3839) plot is 
from Norris and the $V$ magnitudes are taken from Cudworth \& Rees 
(1990) for stars observed in this study with the remainder 
taken from Lee (1977). 
\label{Ivans.fig11}}

\figcaption[Ivans.fig12]{An H-R diagram for the M4 giant stars of the 
present survey.  The symbols are explained in the figure legend, and 
derivation of the $M_{bol}$ values are explained in the text (\S4.1).  
The regions encompassing the RGB and AGB stars (discussed in \S4.2.3) are 
marked on the figure.  We note two stars with wavelength-dependent 
CN-designations.
\label{Ivans.fig12}}

\figcaption[Ivans.fig13]{Carbon abundances (from SS91, adjusted to our 
stellar parameters), nitrogen abundances (determined by syntheses of the CN 
(2--0) bands near \wave{8005}), and C+N+O total abundance  plotted 
versus O abundances (determined by syntheses of the $\lambda\lambda$6300, 
6363${\rm \AA}$ forbidden oxygen lines).  CN-strong and CN-weak stars are 
denoted and the CN strengths appear to correlate with the oxygen abundances.
As earlier, hollow squares ($\Box$) denote values derived for any of the
the three program stars cooler than 3900~K, for which our abundances are 
less reliable.
\label{Ivans.fig13}}

\figcaption[Ivans.fig14]{Sodium abundances (determined by syntheses of the
Na lines at $\lambda\lambda$5682 and 5688~${\rm \AA}$) and aluminum 
abundances (determined by equivalent width analysis of the doublets at 
$\sim\lambda\lambda$6696, 7836, 8773${\rm \AA}$) plotted versus oxygen
abundances.  CN-strong and CN-weak stars are denoted and the CN strengths
appear to correlate with the abundances.  Note that the ordinate and abcissa
are plotted to different scales.  Two stars stand out and are discussed in 
the text (\S 4.2.2).
\label{Ivans.fig14}}

\figcaption[Ivans.fig15]{Sodium abundances and magnesium abundances
(the latter determined by the synthesis of \wave{5711} \nion{Mg}{i} and 
the equivalent width analyses of $\lambda\lambda$5528, 7692${\rm \AA}$) 
plotted versus the aluminum abundances. CN-strong and CN-weak stars are 
denoted and the CN strengths appear to correlate with the Al abundances. 
\label{Ivans.fig15}}

\figcaption[Ivans.fig16]{Sodium abundances plotted versus oxygen
abundances for M4 and M5 (where the scale is the same as that of 
Figure~\ref{Ivans.fig17}). Two stars stand out and are discussed 
in the text (\S 4.2.2 and 5.1.1.).
\label{Ivans.fig16}}

\figcaption[Ivans.fig17]{Sodium abundances plotted versus oxygen
abundances for M4 and other slightly lower-metallicity clusters.
\label{Ivans.fig17}}

\figcaption[Ivans.fig18]{Aluminum abundances plotted versus sodium
abundances for M4, M5, and other slightly lower-metallicity clusters.
\label{Ivans.fig18}}

\figcaption[Ivans.fig19]{Aluminum abundances plotted versus magnesium 
abundances for M4 as well as previous high resolution abundance studies of 
other clusters of a range of metallicities.  In M4, the abundances appear 
to be uncorrelated whereas other clusters show both correlated and 
anticorrelated Mg-Al behavior.  However, as illustrated in the figure, 
the fit between the results we obtained for M4 and those of other clusters 
is gratifying.
\label{Ivans.fig19}}


\begin{references}

\reference{Al75}      Alcaino, G. 1975, \aaps, 21, 5
\reference{AG99a}     Arce, H. G. \& Goodman, A. A. 1999a, \apj, 512. L135
\reference{AG99b}     Arce, H. G. \& Goodman, A. A. 1999b, submitted to 
			\apj, astro-ph/9902110
\reference{ASLK94}    Armosky, B. J., Sneden, C., Langer, G. E., \& 
                        Kraft, R. P. 1994, \aj, 108, 1364
\reference{Bar85}     Barbuy, B. 1985, \aap, 151, 189
\reference{BC77}      Barlow, M. J. \& Cohen, M. 1977 \apj, 213, 737
\reference{Bri97}     Briley, M. M. 1997, \aj, 114, 1051.
\reference{BBHS94}    Briley, M. M., Bell, R. A., Hesser, J. E., \&
                        Smith, G. H. 1994, Can. J. Phys., 72, 772
\reference{BSSBHLS99} Briley, M. M., Smith, V. V., Suntzeff, N. B., Bell,
			R. A., Hesser, J. E., Lambert, D. L., \& Smith, G. H.
			1999, preprint.
\reference{Br98}      Brown, J. A. 1998, private communication
\reference{BW89}      Brown, J. A. \& Wallerstein, G. 1989, \aj, 98, 1643
\reference{BWO90}     Brown, J. A., Wallerstein, G., \& Oke, J. B.
                        1990, \aj, 100, 1561
\reference{BW92}      Brown, J. A. \& Wallerstein, G. 1992, \aj, 104, 1818
\reference{BFBC83}    Buzzoni, A., Fusi Pecci, F., Buonanno, R., \& Corsi,
			C. E. 1983, \aap, 128, 94
\reference{CF83}      Cacciari, C. 1983, \apj, 268, 185
\reference{CS87}      Campbell, B. \& Smith, G. H. 1987, \apj, 323, L69
\reference{CCW78}     Caputo, F., Castellani, V., \& Wood, P. R. 1978,
			\mnras, 184, 377
\reference{CCQ85}     Caputo, F., Castellani, V., \& Quarta, M. L. 1985,
			\aap,143, 8
\reference{CSB98}     Cavallo, R. M, Sweigart, A. V., \& Bell, R. A. 1998,
			\apj, 492, 575.
\reference{Ch81}      Chini, R. 1981, \aap, 99, 346
\reference{CC88}      Clayton, G. C. \& Cardelli, J. A. 1988, \aj, 96, 695
\reference{CFP78}     Cohen, J. G., Frogel, J. A., \& Persson, S. E.
                        1978, \apj, 222, 165
\reference{CD81}      Cottrell, P. L. \& Da Costa, G. S. 1981, \apj, 245, L79
\reference{Cow99}     Cowan, J. J. 1999, private communication
\reference{Cud79}     Cudworth, K. M. 1979, \aj, 84, 1866
\reference{CR90}      Cudworth, K. M. \& Rees, R. F. 1990, \aj, 99, 1491
\reference{dBB85}     de Bi\`evre, P. \& Barnes, I. L. 1985, Internat. J. 
			Mass. Spec. Ion Proc., 65, 211
\reference{dGB91}     de Geus, E. J. \& Burton, W. B. 1991, \aap, 246, 559
\reference{DCCB91}    Dickens, R. J., Croke, B. F. W., Cannon, R. D.,
                        \& Bell, R. A. 1991, \nat, 351, 212
\reference{DL93}      Dixon, R. I. \& Longmore, A. J. 1993, \mnras, 265, 395
\reference{Dor92}     Dorman, B. 1992, \apjs, 81, 221
\reference{DSS92}     Drake, J. J., Smith, V. V., \& Suntzeff, N. B.
                        1992, \apj, 395, L95
\reference{DSS94}     Drake, J. J., Smith, V. V., \& Suntzeff, N. B.
                        1994, \apj, 430, 610
\reference{DSL92}     Dupree, A. K., Sasselov, D. D., \& Lester, J. B. 1992,
			\apj, 387, L85
\reference{EAGLNT93}  Edvardsson, B., Andersen, J., Gustafsson, B., Lambert, 
			D. L., Nissen, P. E., \& Tomkin, J. 1993, \aap, 275, 
			101
\reference{FS87}      Fitzpatrick, M. J. \& Sneden, C. 1987, \baas, 19, 1129
\reference{FPC83}     Frogel, J., Persson, E., \& Cohen, J. 1983, \apjs,  
                        53, 713
\reference{FF97}      Fusi Pecci, F. \& Ferraro, F. R. 1997, private
                        communication
\reference{GL99}      Gay, P. L. \& Lambert, D. L. 1999, private 
			communication
\reference{Gi89}      Gilroy, K. K. 1989, \apj, 347, 835
\reference{GW98}      Gonzalez, G. \& Wallerstein, G. 1998, \aj, in press
\reference{GV88}      Goodrich, R. W. \& Veilleux, S. 1988, \pasp, 100, 1572
\reference{Gra87}     Gratton, R. G. 1987, \aap, 177, 177
\reference{GQO86}     Gratton, R. G., Quarta, M. L., \& Ortolani, S.
                         1986, \aap, 169, 208
\reference{GS91}      Gratton, R. G. \& Sneden, C. 1991, \aap, 241, 501
\reference{GS94}      Gratton, R. G. \& Sneden, C. 1994, \aap, 287, 927
\reference{Gr94}      Gray, D. F. 1994, \pasp, 106, 1248
\reference{GJ91}      Gray, D. F. \& Johanson, H. L. 1991, \pasp, 102, 439
\reference{Gr39}      Greenstein, J. L. 1939, \apj, 90, 397
\reference{Gr68}      Griffin, R. F. 1968, A Photometric Atlas of the Spectrum 
                        of Arcturus $\lambda\lambda3600$--$8825{\rm \AA}$, 
                        Cambridge Philosophical Society
\reference{GBEN75}    Gustafsson, B., Bell, R. A., Ericksson, K., \&
                        Nordlund, A.  1975, \aap, 42, 407
\reference{HSKF98}    Hanson, R. B., Sneden, C., Kraft, R. P., \& Fulbright,
			J. 1998, \aj, 116, 1286
\reference{Har73}     Harris, D. H. 1973 in Interstellar Dust and Related 
                        Topics, IAU Symposium No. 52, ed. J. M. Greenberg 
                        \& H. C. van de Hulst (Dordrecht: Reidel), p. 31
\reference{Hi96}      Hiltgen, D. 1996, Ph.D. Thesis, Univerity of Texas, 
                        Austin
\reference{Ib64}      Iben, I. Jr. 1964, \apj, 140, 1631
\reference{In93}      Ingerson, T. E. 1993, in Fibers Optics in Astronomy II, 
                        ASP Conf. Ser., ed. P. M. Gray, 37, 76
\reference{KB95}      Kemp, S. N. \& Bates, B. 1995, \aaps, 112, 513
\reference{Kr94}      Kraft, R. P. 1994, \pasp, 106, 553
\reference{KSLP92}    Kraft, R. P., Sneden, C., Langer, G. E., \& 
                        Prosser, C. F. 1992, \aj, 104, 645
\reference{KSLS93}    Kraft, R. P., Sneden, C., Langer, G. E., \&
                        Shetrone, M. D. 1993, \aj, 106, 1490
\reference{KSLSB95}   Kraft, R. P., Sneden, C., Langer, G. E., 
                        Shetrone, M. D., \& Bolte, M. 1995, \aj, 109, 2586
\reference{KSSSLP97}  Kraft, R. P., Sneden, C., Smith, G. H., Shetrone, M. D.,
                        Langer, G. E., \& Pilachowski, C. A. 1997, \aj,
                        113, 279
\reference{KSSSF98}   Kraft, R. P., Sneden, C., Smith, G. H., Shetrone, M. D. 
                        \& Fulbright, J. 1998, \aj, 115, 1500
\reference{Kur93}     Kurucz, R. L.1993 in Peculiar versus Normal Phenomena 
			in A-type and Related Stars, ASP Conf. Ser., ed. M. 
			M. Dworetsky, F. Castelli, \& R. Faraggiana, 44, 87
\reference{LM86}      Lambert, D. L. \& McWilliam, A. 1986, \apj, 304, 436
\reference{LMS92}     Lambert, D. L., McWilliam, A. \& Smith, V. V. 1992
			\apj, 386, 685
\reference{LHLD96}    Lambert, D. L., Heath, J. E., Lemke, M., \& Drake, J.
                        1996, \apjs, 103, 183
\reference{LFSB98}    Langer, G. E., Fisher, D., Sneden, C., \& Bolte, M.
                        1998, \aj, 115, 685
\reference{LH95}      Langer, G. E. \& Hoffman, R. 1995, \pasp, 107, 1177
\reference{LHS93}     Langer, G. E., Hoffman, R., \& Sneden, C. 1993,
                        \pasp, 105, 301
\reference{LHZ97}     Langer, G. E., Hoffman, R., \& Zaidins, C. S. 1997,
                        \pasp, 109, 244
\reference{LBS98}     Langer, G. E., Bolte, M., \& Sandquist, E. 1998, 
			\apj, submitted.
\reference{Lee77}     Lee, S.-W. 1977, \aaps, 27, 367
\reference{LJ90}      Liu, T. \& Janes, K. A. 1990, \apj, 360, 561
\reference{LE77}      Lloyd Evans, T. 1977, \mnras, 178, 353
\reference{LBKD95}    Lyons, M. A., Bates, B., Kemp, S. N., \& Davies, R. D.
			 1995, \mnras, 277, 113
\reference{MR94}      McWilliam, A. \& Rich, R. M. 1994, \apjs, 91, 749
\reference{MPSS95}    McWilliam, A., Preston, G. W., Sneden, C., \& Searle, L.
                         1995, \aj, 109, 2757
\reference{MMH66}     Moore, C. E., Minnaert, M. G. J., \& Houtgast, J. 1966
                         The Solar Spectrum 2935{\rm \AA} to 8770{\rm \AA},
                         NBS Mono. 61, (Washington: U.S. Gov. Printing Off.)
\reference{No81}      Norris, J. 1981, \apj, 248, 177
\reference{NB78}      Norris, J. \& Bessell, M. S. 1978, \apj, 225, L49
\reference{NCFD81}    Norris, J., Cottrell, P. L., Freeman, K. C., \& 
			 Da Costa, G. S. 1981, \apj, 244, 105
\reference{ND95a}     Norris, J. E. \& Da Costa, G. S. 1995a, \apj, 441, L81
\reference{ND95b}     Norris, J. E. \& Da Costa, G. S. 1995b, \apj, 447, 680
\reference{NFM96}     Norris, J. E., Freeman, K. C., \& Mighell, K. J. 1996,
			 \apj, 462, 241
\reference{PN89}      Paltoglou, G. \& Norris, J. E. 1989, \apj, 336, 185
\reference{PRC95}     Peterson, R. C., Rees, R. F., \& Cudworth, K. M. 1995
                         \apj, 443, 124
\reference{PSK96}     Pilachowski, C. A., Sneden, C., \& Kraft, R. P.,
                         1996, \aj, 111, 1689
\reference{Roo78}     Rood, R. T. 1972, \apj, 177, 681.
\reference{SFD98}     Schlegel, D. J., Finkbeiner, D. P., \& Davis, M. 1998, 
			\apj, 500, 525
\reference{SH73}      Sawyer Hogg, H. 1973, Publ. David Dunlap Obs, 3, No. 6
\reference{SW75}      Schultz, G. V. \& Wiemar, W. 1975 \aaps, 43, 133
\reference{Sh96a}     Shetrone, M. D. 1996a, \aj, 112, 1517
\reference{Sh96b}     Shetrone, M. D. 1996b, \aj, 112, 2639
\reference{She99}     Shetrone, M. D. 1999, \baas, 30, 1345.
\reference{SK70}      Simoda, M. \& Tanikawa, K. 1970, \pasj, 22, 143
\reference{SN82}      Smith, G. H. \& Norris, J. E. 1982, \apj, 254, 149
\reference{Sd83}      Smith, G. H. \& Dopita, M. A. 1983, \apj, 271, 113
\reference{Smi84}     Smith, G. H. 1984, \aj, 89, 801
\reference{SW91}      Smith, G. H. \& Wirth, G. D. 1991, \pasp, 103, 1158
\reference{SN93}      Smith, G. H. \& Norris, J. E. 1993, \aj, 105, 173
\reference{SSBCB97}   Smith, G. H., Shetrone, M. D., Briley, M. M.,
			Churchill, C. W., \& Bell, R. A. 1997, \pasp,
			109, 236.
\reference{SS89}      Smith, V. V. \& Suntzeff, N. B. 1989, \aj, 97, 1699
\reference{Sn73}      Sneden, C. 1973, \apj, 184, 839
\reference{SNGHYS78}  Sneden, C., Gehrz, R. H., Hackwell, J. A., York, D. G., 
			\& Snow, T. P. 1978, \apj, 223, 168
\reference{SKPL91}    Sneden, C., Kraft, R. P., Prosser, C. F., \&
                        Langer, G. E. 1991, \aj, 102, 2001
\reference{SKPL92}    Sneden, C., Kraft, R. P., Prosser, C. F., \&
                        Langer, G. E. 1992, \aj, 104, 2121
\reference{SKLPS94}   Sneden, C., Kraft, R. P., Langer, G. E., Prosser, C. F.,
                        \& Shetrone, M. D. 1994, \aj, 107, 1773
\reference{SKSSLP97}  Sneden, C., Kraft, R. P., Shetrone, M. D., Smith, G. H.,
                        Langer, G. E., \& Prosser, C. F. 1997, \aj, 114, 1964
\reference{Sun81}     Suntzeff, N. B. 1981, \apjs, 47, 1
\reference{Su93}      Suntzeff, N. B. 1993, in The Globular Cluster-Galaxy
                        Connection, ASP Conf. Ser., ed. G. H. Smith \&
                        J. B. Brodie, 48, 167
\reference{SMTOGW93}  Suntzeff, N. B, Mateo, M., Terndrup, D., Olszewski, 
			E. W, Geisler, D., Weller, W. 1993, \apj, 418, 208
\reference{SS91}      Suntzeff, N. \& Smith, V. 1991, \apj, 381, 160
\reference{Swe97a}    Sweigart, A. V. 1997, \apj, 474, L23
\reference{Swe97b}    Sweigart, A. V. 1997, ``Third Conference on Faint Blue
			Stars'', ed. A. G. D. Philip, J. W. Liebert, \& R.
			A. Safford (Schenectady: L. Davis Press), p. 3
\reference{SW78}      Sweigart, A. V. \& Gross, P. G. 1978, \apjs, 36, 405.
\reference{SM79}      Sweigart, A. V. \& Mengel, J. G. 1979, \apj, 229, 624
\reference{Th90}      Th\'evenin, F. 1990, \aaps, 82, 179
\reference{TL80}      Tomkin, J. \& Lambert, D. L. 1980, \apj, 235, 925
\reference{TELG97}    Tomkin, J., Edvardsson, B., Lambert, D. L., \&
				Gustafsson, B. 1997, \aap, 327, 587
\reference{TMSL95}    Tull, R. G., MacQueen, P. J., Sneden, C., \&
                         Lambert, D. L. 1995, \pasp, 107, 251
\reference{TTC95}     Twarog, B. J., Twarog, B. A., \& Craig, J. 1995, 
			\pasp, 107, 32
\reference{VBM95}     Valenti, J. A., Butler, R. P., \& Marcy, G. W. 1995,
                        \pasp, 107, 966
\reference{Vo87}      Vogt, S. S. 1987, \pasp, 99, 1214
\reference{VCT93}     Vrba, F. J., Coyne S.J., G. V., \& Tapia, S. 1993,
			\aj, 105, 1010.
\reference{Wetal97}   Wallerstein, G, Iben, I. Jr., Parker, P., Boesgaard, 
                        A. M., Hale, G. M., Champagne, A. E.,
                        Barnes, C. A., K{\"a}ppeler, F., Smith, V. V., 
                        Hoffman, R. D., Timmes, F. X., Sneden, C., 
                        Boyd, R. N., Meyer, B. S., \& Lambert, D. L.
                        1997, Rev. Mod. Phys., 69, 995
\reference{WLO87}     Wallerstein, G., Leep, E. M., \& Oke, J. B. 1987,
                        \aj, 93, 1137
\reference{WBBW95}    Whitmer, J. C., Beck-Winchatz, B., Brown, J. A., 
                        \& Wallerstein, G. 1995, \pasp, 107, 127
\reference{Wo94}      Worthey, G. 1994, private communication
\reference{Wou98}     Woudt, P. A. 1998, Ph.D. Thesis, Univerity of Cape
			Town, South Africa.
\reference{ZW84}      Zinn, R. \& West, M. J. 1984, \apjs, 55, 45

\end{references}
\end{document}